\documentclass[useAMS,usenatbib]{mn2e}
\usepackage{graphicx}
\usepackage{amssymb}
\usepackage{color}
\usepackage{hyperref}

\def\be{\begin{equation}} 
\def\ee{\end{equation}}

\def\msun{{\Msun}}

\def\HI{\hbox{H$\scriptstyle\rm I\ $}}

\def\gsim{\lower.5ex\hbox{\gtsima}} 
\def\lsim{\lower.5ex\hbox{\ltsima}} \def\gtsima{$\; \buildrel > \over 
\sim \;$} \def\ltsima{$\; \buildrel < \over \sim \;$} \def\prosima{$\; 
\buildrel \propto \over \sim \;$} \def\gsim{\lower.5ex\hbox{\gtsima}} 
\def\lsim{\lower.5ex\hbox{\ltsima}} 
\def\simgt{\lower.5ex\hbox{\gtsima}} 
\def\simlt{\lower.5ex\hbox{\ltsima}} 
\def\simpr{\lower.5ex\hbox{\prosima}} \def\la{\lsim} \def\ga{\gsim} 
  
 \def\gtsima{$\; \buildrel > \over \sim \;$} 
\def\ltsima{$\; \buildrel < \over \sim \;$} 
\def\gsim{\lower.5ex\hbox{\gtsima}} 
\def\lsim{\lower.5ex\hbox{\ltsima}} 
\def\simgt{\lower.5ex\hbox{\gtsima}} 
\def\simlt{\lower.5ex\hbox{\ltsima}} 
\def\simpr{\lower.5ex\hbox{\prosima}} 
\def\la{\lsim} 
\def\ga{\gsim}

\def\msun{\,{\rm \Msun}}

\def\E3{{\cal E}_{\rm g}^{III}}

\def\Msun{\rm M_\odot}

\def\log{\rm Log}

\def\avchi{$\langle \chi_\mathrm{HI} \rangle$}
 
\title[21cm bispectrum during reionization]{The 21cm bispectrum during reionization: a tracer of the ionization topology}

\author[Hutter et al.]{Anne Hutter$^{1,2}$\thanks{E-mail: a.k.hutter@rug.nl}, Catherine A. Watkinson$^{3}$, Jacob Seiler$^{2,4}$, Pratika Dayal$^{1}$,
\newauthor{Manodeep Sinha$^{2,4}$, Darren J. Croton$^{2,4}$} \\ 
$^{{1}}$ Kapteyn Astronomical Institute, University of Groningen, PO Box 800, 9700 AV Groningen, The Netherlands\\
$^{2}$ ARC Centre of Excellence for All Sky Astrophysics in 3 Dimensions (ASTRO 3D)\\
$^{3}$ Department of Physics, Blackett Laboratory, Imperial College, London, SW7 2AZ, UK \\
$^{4}$ Centre for Astrophysics \& Supercomputing, Swinburne University of Technology, Hawthorn, VIC 3122, Australia
}
\begin{document} 
 
\date{} 
 
%\pagerange{\pageref{firstpage}--\pageref{lastpage}} \pubyear{2014} 
 
\maketitle 
 
\label{firstpage} 
\begin{abstract} 
We compute the bispectra of the 21cm signal during the Epoch of Reionization for three different reionization scenarios that are based on a dark matter N-body simulation combined with a self-consistent, semi-numerical model of galaxy evolution and reionization. Our reionization scenarios differ in their trends of ionizing escape fractions ($f_\mathrm{esc}$) with the underlying galaxy properties and cover the physically plausible range, i.e. $f_\mathrm{esc}$ effectively decreasing, being constant, or increasing with halo mass.
We find the 21cm bispectrum to be sensitive to the resulting ionization topologies that significantly differ in their size distribution of ionized and neutral regions throughout reionization. From squeezed to stretched triangles, the 21cm bispectra features a change of sign from negative to positive values, with ionized and neutral regions representing below-average and above-average concentrations contributing negatively and positively, respectively.
The position of the change of sign provides a tracer of the size distribution of the ionized and neutral regions, and allows us to identify three major regimes that the 21cm bispectrum undergoes during reionization. In particular the regime during the early stages of reionization, where the 21cm bispectrum tracks the peak of the size distribution of the ionized regions, provides exciting prospects for pinning down reionization with the forthcoming Square Kilometre Array.
\end{abstract}

 
% ******************************************************************************************************
\section{Introduction}
% ******************************************************************************************************

The Epoch of Reionization (EoR) marks a major phase transition in the history of the Universe. High-energy photons from the first stars and galaxies permeate the intergalactic medium (IGM) and gradually ionize the neutral hydrogen (\HI) of the IGM until the Universe is fully ionized at $z\simeq6$ \citep{Fan2006, Dayal2018}. While the number of constraints on the timing of reionization has been increasing in the past years \citep{Planck2018, Mortlock2011, Bolton2011, Ouchi2018, Konno2018, Pentericci2014, Schenker2014, Hutter2014, Hutter2015}, the exact time evolution of the ionization fraction and the percolation of the ionized regions into the IGM remain highly uncertain. 

With the first galaxies being amongst the key sources of hydrogen reionization, the time and spatial evolution of the ionized regions (the ionization topology) will naturally be strongly linked to the properties of the underlying galaxies and their location in the IGM. A key property of these high-redshift galaxies is the escape fraction of hydrogen ionizing photons ($f_\mathrm{esc}$) which describes the fraction of ionizing photons that escape from the galactic environment into the IGM. This escape fraction has been shown to be highly dependent on the physical processes and the resulting gas distribution within and around galaxies \citep{Paardekooper2015, Kimm2017, Kimm2019, Seiler2018, Trebitsch2018}. 
On the one hand, outflows generated by radiative feedback \citep[e.g.][]{Kitayama2004, Whalen2004, Abel2007}, supernovae explosions \citep{Kimm2017, Kimm2019} or mass accretion onto black holes \citep{Seiler2018, Trebitsch2018} can generate low density paths through which the ionizing radiation can escape into the IGM. On the other hand, as a galaxy accretes more mass over time, its gravitational potentials deepens, and the susceptibility of its gas being ejected by these processes decreases. This allows two competing scenarios: one where the escape fraction decreases with halo mass, assuming that the deepening of the gravitational potential dominates either/both the star formation rate and mass accretion providing the outflow energy \citep[e.g.][]{Kimm2017, Kimm2019}. The other is where the escape fraction increases with halo mass, presuming that the increase in star formation rate and porosity of the interstellar medium (ISM) with halo mass creates more escape paths for the ionizing radiation \citep{Wise2009}. These contrasting dependencies of the escape fraction with mass can result in ionization topologies which are quite different. 

Detections of the 21cm signal, from neutral hydrogen in the early Universe, with current and future radio interferometers, such as the Low Frequency Array \citep{vanHaarlem2013}, the Murchison Widefield Array \citep{Tingay2013} and the Square Kilometre Array \citep{Carilli2004}, will allow us to measure the spatial distribution of the ionized regions throughout reionization and hence provide us with constraints on high-redshift galaxy properties. Indeed, several theoretical works have found a dependency between the properties of the ionizing sources and the 21cm signal \citep[e.g.][]{McQuinn2007, Iliev2012, Kim2013, Geil2016, Seiler2019}. In particular, \citet{Seiler2019} and similarly \citet{Kim2013} have shown that measurements of the 21cm power spectrum will be able to constrain the trends of the dependency of $f_\mathrm{esc}$ with halo or stellar mass. 

However, as of now, these different scenarios have only been studied with the Gaussian part of the signal, which ignores key non-Gaussian information available in the 21cm maps. Recently, these non-Gaussianities of the 21cm signal have been analysed using two statistical approaches: (i) 3-point correlation functions \citep{Hoffmann2018, Gorce2019}, which can provide good tracers of the characteristic scale of the ionized and neutral regions during the early and late stages of reionization, respectively.
(ii) The 21cm bispectrum \citep{Bharadwaj2005, Shimabukuro2016, Majumdar2018, Watkinson2019}, which has been found to follow the matter and spin temperature fluctuations during Cosmic Dawn \citep{Watkinson2019} and the early stages of reionization \citep{Majumdar2018}, with its amplitude being sensitive to the distribution and emissivity of X-ray sources. This is because, as X-ray sources become more luminous and rare, the imprint of their heating profile shape becomes more pronounced \citep{Watkinson2019}. During reionization the 21cm bispectrum is mostly governed by the non-Gaussianities of the neutral hydrogen fraction fluctuations \citep{Shimabukuro2016, Majumdar2018}. Importantly, \citet{Majumdar2018} find that the sign of the 21cm bispectrum changes depending on whether its non-Gaussianities are driven by the fluctuations in the neutral fraction field (negative) or fluctuations in the matter density field (positive). While these findings highlight the potential of the 21cm bispectrum and its power to tighten constraints on EoR source models \citep{Shimabukuro2017, Majumdar2018}, the physical link between characteristic features in the 21cm bispectrum, the ionization topology, and the escape fraction of ionizing photons remains unclear as of now. 

This paper aims at bridging this gap by addressing the following: What are the signatures of an inside-out reionization topology, i.e. do ionization fronts percolate from the over- to under-dense regions? Is the ionized bubble size distribution imprinted in the 21cm bispectra? How sensitive is the 21cm bispectra to different ionization topologies? How does the 21cm bispectra evolve throughout reionization? To answer these questions, we analyse the 21cm bispectra of the reionization simulations described in \citet{Seiler2019} that cover the physically plausible parameter space by exploring three models where the escape fraction either remains constant, decreases or increases with halo mass. The analysis of the resulting different ionization topologies allows us to find a descriptive physical interpretation of the 21cm bispectrum characteristics.

This paper is organised as follows. In Section \ref{sec_model} we describe our semi-numerical reionization simulations. Section \ref{sec_computing_bispectra} includes our modelling of the 21cm signal and the computation of the bispectra. We analyse the bispectra of the neutral hydrogen fraction fluctuations and their dependencies on the assumed $f_\mathrm{esc}$ model in Section \ref{sec_HI_bispectra}. In Section \ref{sec_21cm_bispectra} we extend our analysis to the 21cm bispectra. We then conclude in Section \ref{sec_conclusions}. Throughout this paper we assume a $\Lambda$CDM Universe with cosmological parameter values of $\Omega_\Lambda=0.698$, $\Omega_m = 0.302$, $H_0=100h=68.1$km~s$^{-1}$Mpc$^{-1}$ and $\sigma_8=0.828$.

% **************************************************************************************************
\section{Reionization simulations}
\label{sec_model}
% **************************************************************************************************

In this Section we briefly describe our self-consistent, semi-numerical reionization simulations performed with {\sc rsage}\footnote{\url{https://github.com/jacobseiler/rsage}} \citep{Seiler2019}.  
We refer the interested reader to \citet{Seiler2019} for details.

\subsection{N-body simulation}

Our reionization simulations are based on the dark matter N-body simulation {\it Kali} that was run with the SPH-code {\sc gadget-3} \citep{Springel2005}. {\it Kali} contains $2400^3$ dark matter particles in a simulation box with a side length of $160$~Mpc. Gravitationally bound structures of at least $32$ particles are identified as halos using {\sc subfind} \citep{Springel2001}, resulting in a minimum halo mass of $\sim4\times10^8\msun$. Snapshots of the particles are stored every $10$~Myrs, between $z=30$ and $z=5.5$, resulting in a total of $98$ snapshots. In order to follow the evolution of galaxies and reionization, merger trees have been built using {\sc gbptrees} \citep{Poole2017} and the dark matter distributions within the simulation box have been mapped to $1024^3$ and down-sampled to $256^3$ grids for each snapshot that is output. These provide the required input data for running {\sc rsage}.

\subsection{RSAGE}

{\sc rsage} couples the semi-numerical reionization code {\sc cifog}\footnote{\url{https://github.com/annehutter/grid-model}} \citep{Hutter2018a} to the semi-analytic galaxy formation model {\sc sage}\footnote{\url{https://github.com/darrencroton/sage}} \citep{Croton2016}. 
For each galaxy {\sc sage} follows gas accretion from the IGM, metal-dependent gas cooling, star formation, metal enrichment of the gas, gas heating and outflows due to supernovae explosions and AGNs, and mergers. {\sc rsage} includes some modifications to the {\sc sage} model: Firstly, due to the shorter dynamical time scales at high redshifts $z\gtrsim7$, high-mass stars are not assumed to instantaneously explode as supernovae as commonly done in galaxy formation models at lower redshifts \citep[e.g.][]{Croton2016} but with delays that correspond to their life times sampled by the initial mass function \citep[e.g.][]{Mutch2016}. Secondly, by computing the evolution of the ionized regions around galaxies during reionization and the corresponding photoionization rates, {\sc rsage} also accounts for the radiative feedback from local reionization on the gas content in each galaxy.

In order to facilitate the coupling between the high-redshift version of {\sc sage} and {\sc cifog}, {\sc rsage} computes the number of ionizing photons from the stellar mass history of each galaxy. It uses the age and metallicity dependent ionizing photon yields from the stellar population synthesis code {\sc starburst99} \citep{Leitherer1999}. The resulting number of ionizing photons is mapped to a grid at the galaxy's location and fed into {\sc cifog}. From the fields, containing the cumulative number of ionizing photons and gas density, {\sc cifog} computes the distribution of the ionized regions as well as the spatially-dependent photoionization rates. In doing so, it also accounts for recombinations and tracks the residual \HI fraction in ionized regions, both depending on the local photoionization rate and gas density. From the resulting photoionization rates at the galaxies' locations, the suppression of baryonic infall due to radiative feedback is calculated following the critical mass relations found in equation 3 in \citet{Sobacchi2013}.

\subsection{Ionizing escape fraction models}

In our reionization simulations, the galaxy evolution model parameters in {\sc rsage} have been tuned to reproduce the observed stellar mass functions at $z=8$, $7$ and $6$. With the galaxy evolution model parameters being fixed, we perform three different reionization simulations, where the escape fraction of ionizing photons $f_\mathrm{esc}$ is varied. In two of our simulation runs the ionizing escape fraction is coupled to self-consistently computed galaxy properties, while the third run serves as a benchmark and assumes an overall constant ionizing escape fraction. Our $f_\mathrm{esc}$ models therefore are:
\begin{enumerate}
 \item {\bf Constant:} The ionizing escape fraction $f_\mathrm{esc}$ is assumed to have a constant value of $20$\% for all galaxies at all times.
 \item {\bf Ejected:} The ionizing escape fraction $f_\mathrm{esc}$ depends on the fraction of gas expelled by supernovae and quasar feedback from the galaxy, $f_\mathrm{ej}$, 
 \begin{eqnarray}
  f_\mathrm{esc} &=& \alpha\ f_\mathrm{ej} + \beta,
 \end{eqnarray}
 with $\alpha=0.3$ and $\beta=0$.
 This model follows the findings of radiation-hydrodynamical simulations where the escape of ionizing photons is enhanced by low-density tunnels created by supernovae explosions \citep{Kimm2017, Paardekooper2015}. For an average galaxy, this model results in $f_\mathrm{esc}$ decreasing with halo mass.
 \item {\bf SFR:} The ionizing escape fraction $f_\mathrm{esc}$ scales with the star formation rate of the galaxy as
 \begin{eqnarray}
  f_\mathrm{esc} &=& \frac{\delta}{1 + \exp(-\alpha\ [~\log_{10}(\mathrm{SFR}) - \beta~])},
 \end{eqnarray}
 with $\alpha=1$, $\beta=1.5$ and $\delta=1$. The logistic curve was chosen as its argument $\log_{10}(\mathrm{SFR})$ can span $[-\infty, +\infty]$ while its value ranges between $0$ and $\delta$. Since the $f_\mathrm{esc}$ values can not exceed unity, we set $\delta=1$. Furthermore, $\beta$ depicts the value of $\log_{10}(\mathrm{SFR})$ that corresponds to $\delta/2$. For an average galaxy, this model results in $f_\mathrm{esc}$ increasing with halo mass.
\end{enumerate}

All $f_\mathrm{esc}$ prescriptions have been tuned to reproduce the optical depth measurements by \citet{Planck2018}, i.e. $\tau\simeq0.055$, and the evolution of the ionizing emissivity derived from high-redshift galaxy observations \citep{Bouwens2015}.

% ***************************************************************************************************
\section{Computing the HI \& 21cm bispectra}
\label{sec_computing_bispectra}
% ***************************************************************************************************

\subsection{The 21cm signal}

The \HI 21cm line corresponds to the transition between the triplet and singlet hyperfine state of the hydrogen atom in its ground state. During Cosmic Dawn and the Epoch of Reionization this line can be seen either in absorption or emission against the cosmic microwave background (CMB). Hence, the measurable 21cm signal is given by the difference between the attenuated intensity of the CMB and any induced 21cm emission following the absorption of CMB photons. Both processes are  sensitive to the distribution of the hydrogen atoms in the singlet and triplet hyperfine states described by the spin temperature $T_s$. Assuming that the spin temperature ($T_s$) is well heated above the CMB temperature ($T_\mathrm{CMB}$), the measurable differential 21cm brightness temperature is given by \citep[e.g.][]{Iliev2012}
\begin{eqnarray}
 \delta T_b (\bf{x}) &=& T_0\ \left[1+\delta(\bf{x})\right]\ \chi_\mathrm{HI}(\bf{x}), \\
 T_0 &=& 28.5 \mathrm{mK}\ \left(\frac{1+z}{10}\right)^{1/2} \frac{\Omega_b}{0.042} \frac{h}{0.073} \left(\frac{\Omega_m}{0.24}\right)^{-1/2}, \nonumber
\end{eqnarray}
with $\delta({\bf x})$ being the gas over-density $\rho({\bf x})/\overline{\rho}$ at position ${\bf x}$, $\overline{\rho}$ the mean gas density, and $\chi_\mathrm{HI}$ the neutral hydrogen fraction. We note that this assumption could break at the very early stages of reionization ($\chi_\mathrm{HI}\lesssim0.1$) and the 21cm signal could be overestimated, particularly in the under-dense regions far from galaxies where the IGM (and thereby $T_s$) has not been sufficiently heated above $T_\mathrm{CMB}$.
We compute the differential 21cm brightness temperature fields from the ionization ($\chi_\mathrm{HII}$) and density ($\rho/\overline{\rho}$) fields ($256^3$ grids) for all available snapshots between the emergence of the first ionizing sources at $z\simeq15$ and the completion of reionization at $z\simeq6$.

\subsection{Bispectrum}

The bispectrum of a field $T(\bf{x})$ is defined as the Fourier transform of the three-point correlation function. For statistically homogeneous and isotropic fields the bispectrum $B$ remains unchanged under translations and rotations, and is described by,
\begin{equation}
\scalebox{0.9}{$
 (2\pi)^3 B(k_1, k_2, k_3)\ \delta_\mathrm{D} (\bf{k_1}+\bf{k_2}+\bf{k_3}) = \langle \Delta({\bf k_1}) \Delta({\bf k_2}) \Delta({\bf k_3}) \rangle,
 $}
 \label{eq_bispectrum}
\end{equation}
where $\Delta({\bf k})$ is the Fourier transform of $T(\bf{x})$, and $\delta_\mathrm{D}$ is the delta function, which yields $1$ if $\bf{k_1}+\bf{k_2}+\bf{k_3}=0$ and $0$ otherwise. This implies that only closed triangles in $k$-space contribute to the bispectrum. While the three-point correlation function describes the excess probability of a spatial configuration of three points in real space, the bispectrum provides a measurement to which degree a structure described by the closed triangle in $k$-space is present. 

In contrast to the Fourier transformation of the two-point correlation function, the power spectrum, the bispectrum can be positive or negative. The sign of the bispectrum is sensitive to whether the structure given by the $(\bf{k_1}, \bf{k_2}, \bf{k_3})$ triangle in $k$-space coincides with the shape of above- or below-average concentrations in the field $T({\bf x})$. 
Above-average concentrations or enhanced over-densities contribute positively to the bispectrum, while below-average concentrations or enhanced under-densities contribute negatively. Hence, the sign of the bispectrum indicates whether above- or below-average concentrations predominate.
The closer the concentrations of the above- or below-average signal are to a given triangle configuration's interference pattern, the stronger the bispectrum will be for that configuration relative to others. Such structure also contributes to the bispectrum of other triangle configurations non-negligibly.

\begin{figure}
 \centering
 \includegraphics[width=0.49\textwidth]{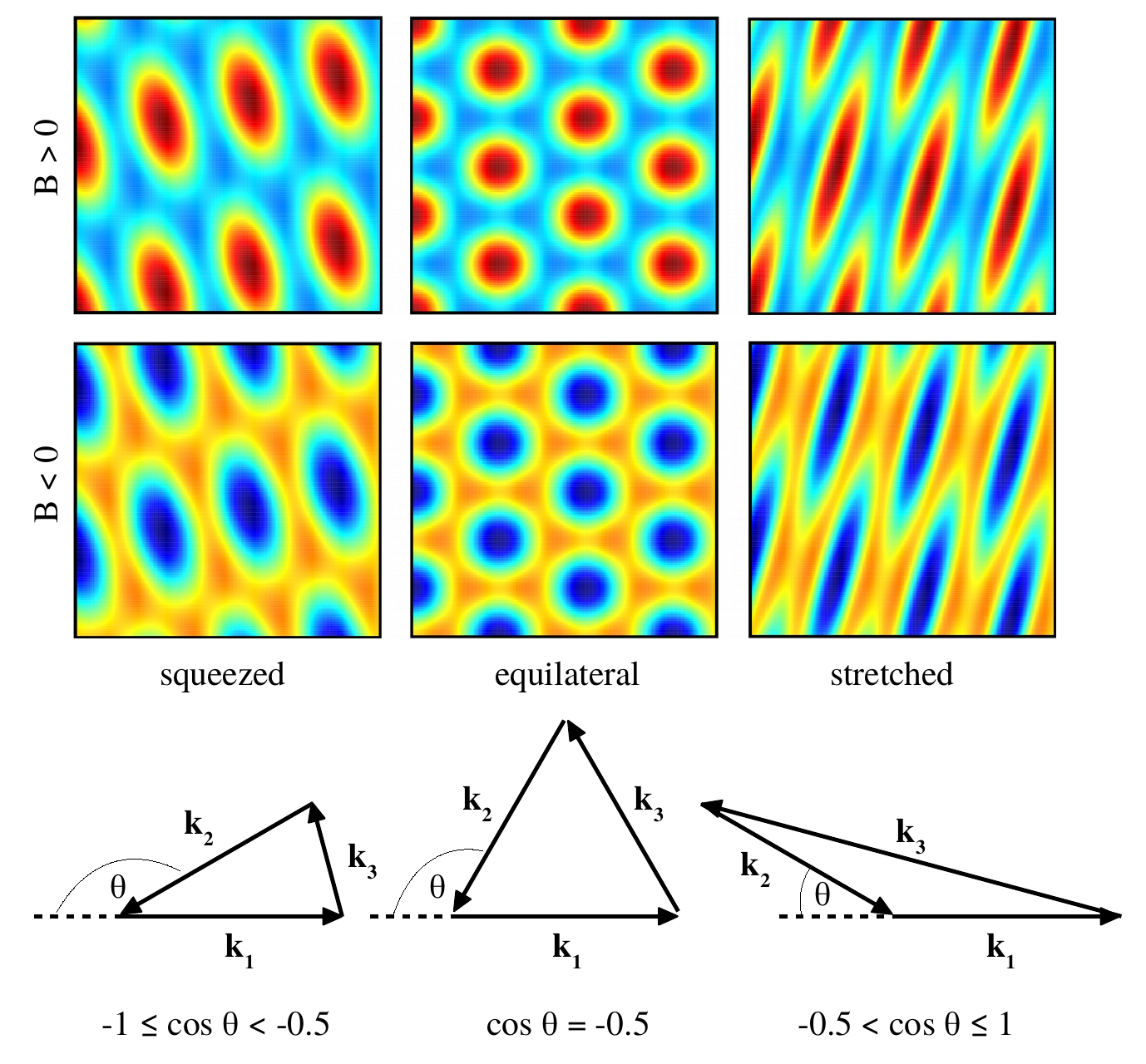}
 \caption{Different regimes of closed triangle configuration in $k$-space (bottom) and their corresponding real space fluctuations (top). A positive bispectrum indicates above-average concentrations (red regions), and a negative bispectrum indicates below-average concentrations (blue regions). The schematics also depict our definition of the angle $\theta$ that is used throughout this paper.}
 \label{fig_triangle_configurations}
\end{figure}

Fig. \ref{fig_triangle_configurations} illustrates the different regimes of closed triangle configurations and their corresponding fluctuations in real space projected on a 2D plane. The 3D interference pattern of the three $k$ vectors forming a closed triangle extends perpendicular to the depicted plane - therefore Fig. \ref{fig_triangle_configurations} shows the cross section of the resulting filaments. 
As can be seen from the middle panel, the equilateral configuration ($\cos\theta=-0.5$) represents triangles whose real space fluctuations are closest to filaments with a circular cross section. As we shorten one leg of the triangle in $k$-space, i.e. moving from the equilateral to the squeezed triangles, the cross section of the filaments in real space becomes more elongated. Likewise the filamentary structures become more plane-like, as we move towards stretched triangles, i.e. stretching one leg of the triangle in $k$-space. While the trends towards $\cos\theta=-1$ and $\cos\theta=1$ seem similar, their limiting cases are rather different. A fully stretched ($\cos\theta=1$) triangle corresponds to plane-like fluctuations, while a fully squeezed ($\cos\theta=-1$) triangle describes plane-like structures with a large-scale modulation given by the shortest leg ($k_3$ in Fig. \ref{fig_triangle_configurations}). For fully squeezed triangles a positive (negative) bispectrum indicates that there is more (less) small-scale structure where there is a large-scale above-average concentration, and less (more) small-scale structure where there is a large-scale below-average concentration \citep[see also][]{Lewis2011}.

From our ionization and density fields we derive the $\chi_\mathrm{HI}$ and 21cm bispectra using the FFT-bispectrum estimator\footnote{Our MPI-parallelised implementation of the FFT polyspectrum estimator is available at \url{https://github.com/annehutter/polyspectrum}. We have checked that it reproduces the analytic test cases shown in Fig. 7 in \citet{Watkinson2017}.} 
described in \citet{Watkinson2017}. This algorithm recasts equation \ref{eq_bispectrum} and utilises a number of fast Fourier transformations. In contrast to direct bispectrum measurements where the triangles are explicitly constructed, the Dirac delta function in equation \ref{eq_bispectrum} is enforced by using filters in k-space.
For a more detailed description of the algorithm we refer the reader to \citet{Watkinson2017} and comment here only on our choice of binning. We aim to reduce the statistical noise by binning the bispectrum over $\cos\theta\pm0.05$, with $\theta$ being the angle between the triangle legs $k_1$ and $k_2$. Our bispectrum calculation includes constructing a filter for all $k_1$, $k_2$ and $k_3$ values in k-space. The discrete nature of our ionization fields leads to all $k$-values having an uncertainty that corresponds to at least the width of a cell in $k$-space. While we could increase this uncertainty in $k$ to increase the number of triangles probed and thus reduce the noise in our statistics, we refrain from doing so. This is because such a measure would result in smoothing over an unreasonable range of scales, especially at larger real-space scale lengths, and lead to the disappearance of otherwise clear features. We show the number of triangles probed in all bispectra shown in this paper in Fig. \ref{fig_Ntriangles} and estimate the uncertainties due to statistical fluctuations in Appendix \ref{app_stats_fluctuations}.

Throughout this paper, we do not analyse the raw bispectrum $B(k_1, k_2, k_3)$ (see equation \ref{eq_bispectrum}) but the normalised bispectrum defined as \citep{Watkinson2017}, 
 \begin{eqnarray}
  \tilde{B}(k_1, k_2, k_3) &=& \frac{B(k_1, k_2, k_3)}{\sqrt{k_1\ k_2\ k_3\ P(k_1)\ P(k_2)\ P(k_3)}}.
 \end{eqnarray}
This normalisation has two advantages: firstly, the 21cm bispectrum becomes dimensionless, and secondly it isolates non-Gaussianities by normalising out the contribution of the respective power spectra. We refer the reader to \citet{Watkinson2019} for a study of the different normalisations of the 21cm bispectrum and their applications.

\begin{figure}
 \includegraphics[width=0.45\textwidth]{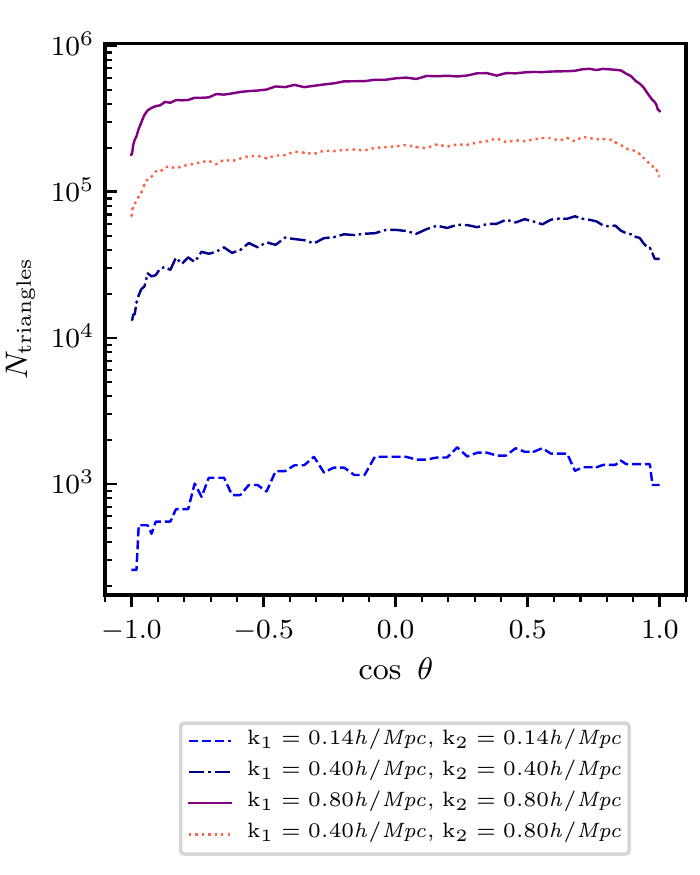}
 \caption{Number of triangles for all triangle configurations of the bispectra shown in Fig. \ref{fig_dimlessbispectra_XHI_multiple} and \ref{fig_dimlessbispectra_21cm_multiple}.}
 \label{fig_Ntriangles}
\end{figure}

% ***************************************************************************************************
\section{Bispectra of the $\chi_\mathrm{HI}$ fluctuations}
\label{sec_HI_bispectra}
% ***************************************************************************************************

In order to understand the characteristics of the 21cm signal bispectra, we first analyse and discuss the bispectra of the \HI fractions, $\chi_\mathrm{HI}(\bf{x})$, throughout reionization. The different ionization topologies of our reionization simulations, arising from different $f_\mathrm{esc}$ descriptions, provide us with the unique opportunity to identify their common and distinguishing characteristics.

\begin{figure*}
 \includegraphics[width=0.8\textwidth]{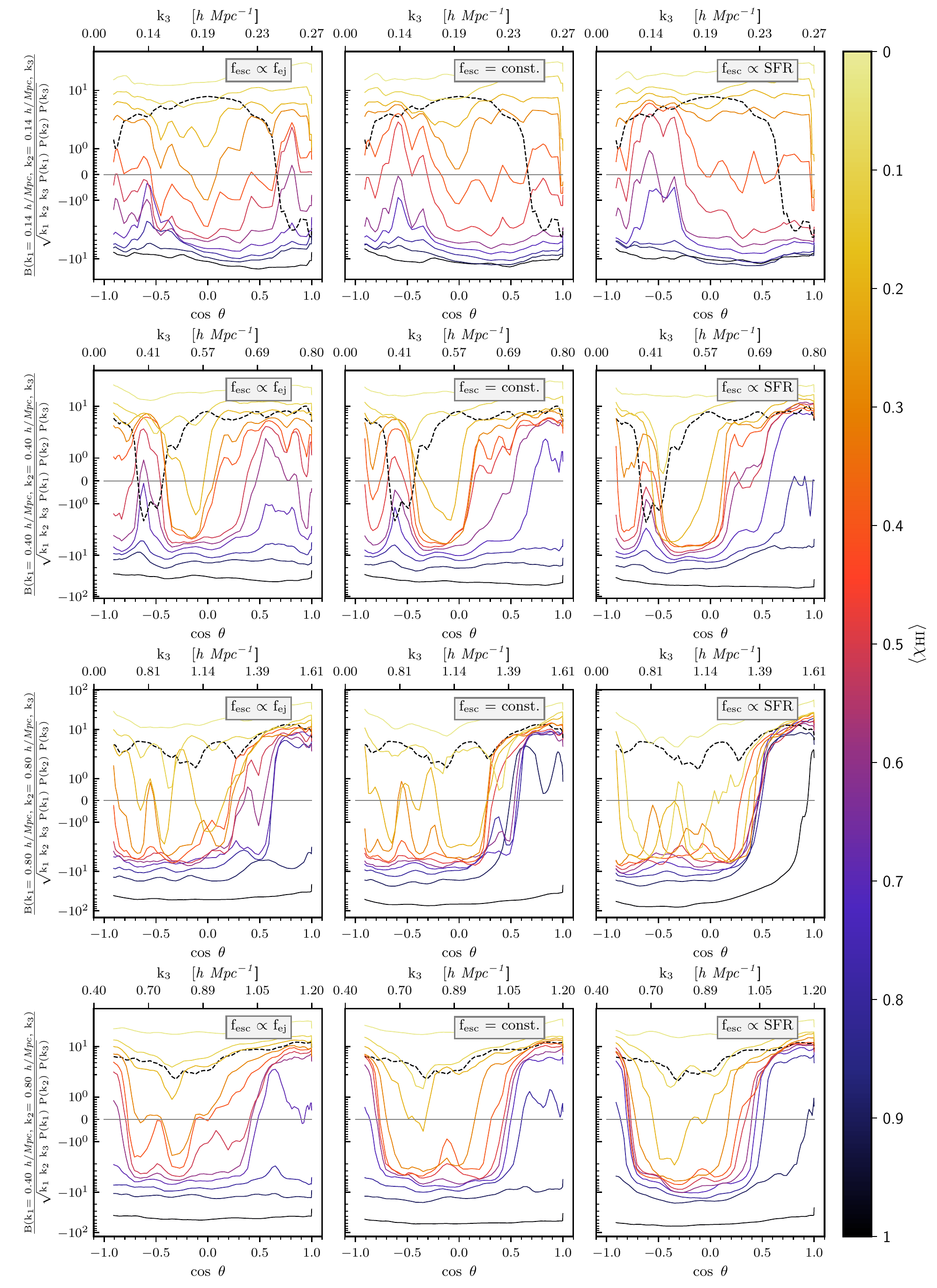}
 \caption{Normalised bispectra of the neutral fraction fluctuations at $\langle \chi_\mathrm{HI} \rangle=0.02$, $0.1$, $0.2$, $0.3$, $0.4$, $0.5$, $0.6$, $0.7$, $0.8$, $0.9$, $0.99$ as indicated by the coloured lines. The black dashed line indicates the normalised bispectrum of the underlying dark matter density field at $\langle \chi_\mathrm{HI} \rangle=0.5$. The left, centre and right columns show the normalised bispectra for the $f_\mathrm{esc}$ models that decrease, remain constant, and increase with halo mass, respectively. The first three rows show the normalised bispectra for isosceles triangles with $k_1=k_2=0.14h$~Mpc$^{-1}$ (probing large scales of $\sim20h^{-1}$Mpc), $0.4h$~Mpc$^{-1}$ (probing intermediate scales of $\sim8h^{-1}$Mpc), $0.8h$~Mpc$^{-1}$ (probing small scales of $\sim4h^{-1}$Mpc), and the last row the normalised bispectra for non-isosceles triangles with $k_1=\frac{1}{2}k_2=0.4h$~Mpc$^{-1}$.}
 \label{fig_dimlessbispectra_XHI_multiple}
\end{figure*}

\subsection{Global Trend}

Fig. \ref{fig_dimlessbispectra_XHI_multiple} shows the normalised $\chi_\mathrm{HI}$ bispectra of our three models for {\it isosceles} ($k_1=k_2\neq k_3$, upper three rows) and {\it non-isosceles} triangles ($k_1=\frac{1}{2} k_2\neq k_3$, bottom row) as reionization proceeds. We can see that the $\chi_\mathrm{HI}$ bispectrum shows the same global qualitative trends for all models and triangle configurations: during the first half of reionization the normalised bispectrum is negative, with the negative amplitude decreasing as reionization approaches its midpoint at \avchi$\simeq0.5$. As reionization proceeds further, the bispectrum becomes positive and its amplitude increases as the neutral hydrogen content drops.
This trend can be understood as follows: before reionization reaches its midpoint, the majority of the volume is neutral and ionized regions represent below-average $\chi_\mathrm{HI}$ concentrations. This abundance of below-average concentrations results in a negative bispectrum. As the Universe becomes increasingly ionized, the difference between the $\chi_\mathrm{HI}$ values in the ionized regions and \avchi~decreases, i.e. the contrast drops and so does the negative amplitude of the bispectrum. As reionization passes its midpoint, the majority of the volume becomes ionized and neutral regions become more concentrated. These above-average concentrations in terms of $\chi_\mathrm{HI}$ cause the bispectrum to become positive. Similarly, as \avchi~decreases further the contrast between the neutral regions and \avchi~increases, leading to an increase in the amplitude of the bispectrum.

This trend is well seen for bispectra probing large-scale fluctuations (c.f. $k_1=k_2=0.14h$~Mpc$^{-1}$, first row in Fig. \ref{fig_dimlessbispectra_XHI_multiple}). However, as we probe smaller scales, the size and shapes of ionized and neutral regions start to dominate the $\chi_\mathrm{HI}$ bispectra, leading to a mostly negative bispectrum that becomes positive when it tracks the neutral plane-like structures as we discuss in Section \ref{subsec_ion_and_neutral_sizes}.

\subsection{Large scale reionization topology}

Being sensitive to the sizes and shapes of the ionized regions in the IGM, the relation between the $\chi_\mathrm{HI}$ bispectra and density bispectra can reveal whether the ionization fronts percolate from over-densities, where the galaxies are located, to under-dense voids (inside-out) or vice versa (outside-in). In an inside-out scenario, the ionized regions that are forming in over-dense regions represent below-average concentrations in the $\chi_\mathrm{HI}$ field, while the same regions correspond to above-average concentration in the density field; hence, the $\chi_\mathrm{HI}$ bispectra and density bispectra should show the opposite trends on larger scales. 

Indeed, since our three reionization simulations feature inside-out topologies, we find the maxima in the $\chi_\mathrm{HI}$ bispectra to correspond to the minima in the density bispectrum for large-scale modes, i.e. $k\lesssim 0.4 h$~Mpc$^{-1}$. In Fig. \ref{fig_dimlessbispectra_XHI_multiple} this is prominently seen for the most sphere-like shapes with $k_1=k_2=0.4h$~Mpc$^{-1}$ at scales of $k_3\simeq0.36h$~Mpc$^{-1}$, corresponding to sizes of $r=\pi/k_3\simeq9h^{-1}$~Mpc. As can be seen from the black dashed line in all panels of the second row in Fig. \ref{fig_dimlessbispectra_XHI_multiple}, showing the normalised bispectrum of the underlying density field at \avchi$\simeq0.5$, voids in our simulations lead to a negative minimum in the density bispectrum at $k_3\simeq0.36h$~Mpc$^{-1}$. At the same scale, the absence of ionized regions in voids causes the $\chi_\mathrm{HI}$ bispectrum to be maximal in amplitude (c.f. dark purple to orange solid lines in second row in Fig. \ref{fig_dimlessbispectra_XHI_multiple}). 

Naturally, the amplitude of this maximum in the $\chi_\mathrm{HI}$ bispectrum depends on the exact ionization topology, in particular by how much the ionized regions deviates from the underlying density fields. Hence, increasing the bias of the ionizing emissivity, i.e. going from the {\it ejected} ($f_\mathrm{ej}$ in our plot labels) to the {\it SFR} $f_\mathrm{esc}$ model, results in voids being more ``confined'' or reduced in size by the surrounding ionized regions. This decrease in the anti-correlation between the density and $\chi_\mathrm{HI}$ fields leads to a less pronounced maximum in the $\chi_\mathrm{HI}$ bispectrum. For example, while the maximal amplitude of the normalised $\chi_\mathrm{HI}$ bispectrum at \avchi$=0.6$ is $\tilde{B}\sim1$ for the {\it ejected} $f_\mathrm{esc}$ model ($f_\mathrm{esc}\propto f_\mathrm{ej}$), it drops down to $\tilde{B}\sim -0.8$ for the {\it SFR} $f_\mathrm{esc}$ model ($f_\mathrm{esc}\propto \mathrm{SFR}$).

\begin{figure*}
 \includegraphics[width=0.9\textwidth]{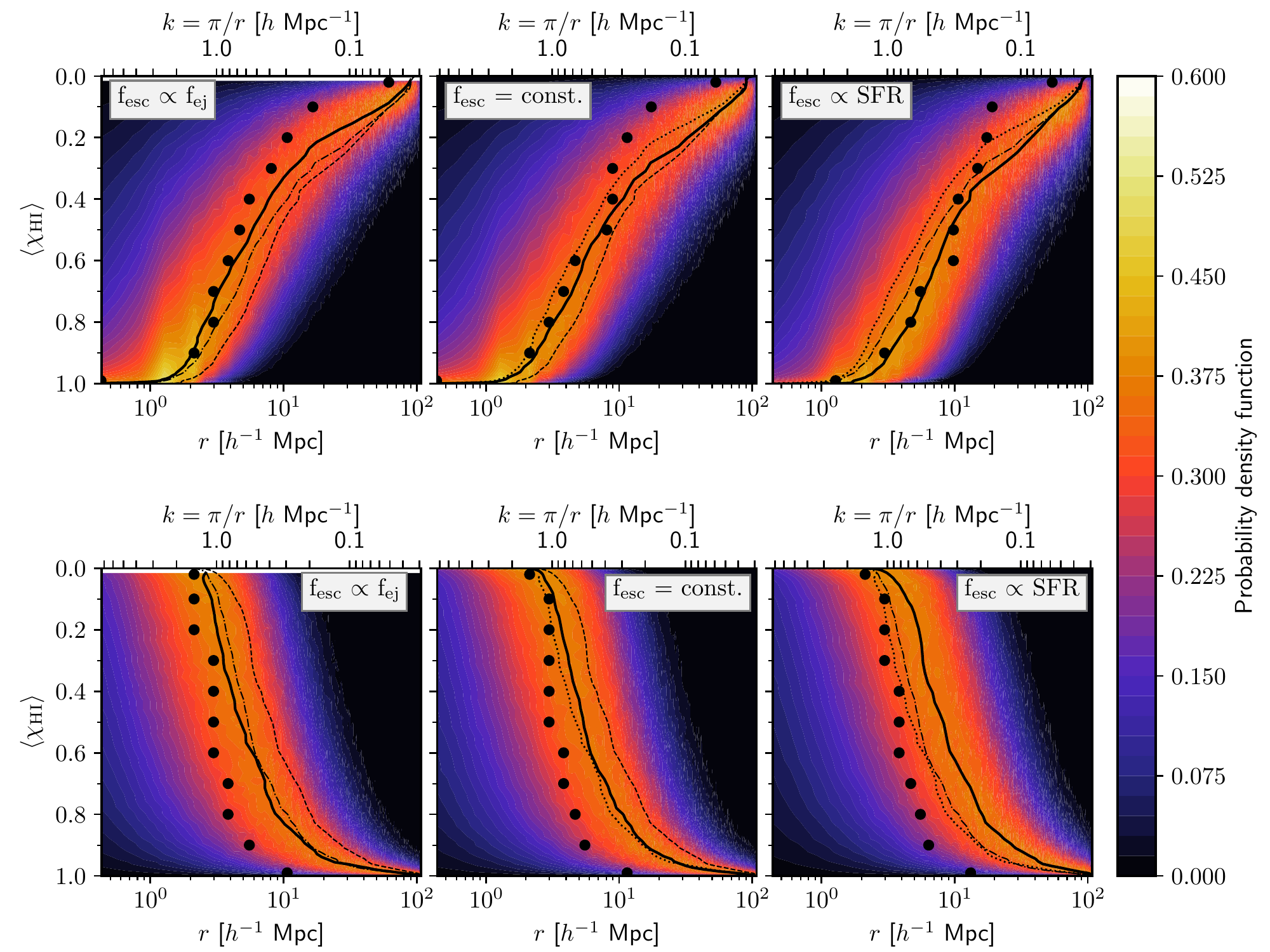}
 \caption{Top: Probability density distribution (PDF) for the sizes of the ionized regions. Bottom: PDF for the sizes of the neutral regions. In each panel, colour encodes the amplitude of the PDF, while the solid black line shows the evolution of the maximum of the PDF throughout reionization. Black dots mark the maxima of the respective granulometry size distribution shown in Fig. \ref{fig_ion_neutral_granulometry}. For comparison, we plot the maxima of the PDFs of the other two $f_\mathrm{esc}$ models in each panel, i.e. thin dotted, dash-dotted and dashed lines correspond to the {\it ejected}, {\it constant} and {\it SFR} $f_\mathrm{esc}$ models, respectively.}
 \label{fig_ion_neutral_bubbledistr}
\end{figure*}

\subsection{Shapes and sizes of ionized \& neutral regions}
\label{subsec_ion_and_neutral_sizes}

In order to link the characteristic features in the $\chi_\mathrm{HI}$ bispectrum with the ionization topology, we start by discussing the evolution of the sizes and shapes of the ionized and neutral regions during reionization.
In the beginning of reionization galaxies ionize bubbles around themselves. As long as individual ionized regions do not overlap, the size of each ionized region depends mostly on the ionizing emissivity of the underlying galaxy. Due to the isotropy of the ionizing radiation emitted from galaxies, the shape of the ionized regions are most likely to resemble spheres. 
However, as the ionized regions start to merge, their shapes become more complex and their size distribution broadens, comprising scales from single to multiple merged bubbles. This broadening of the size distribution of the ionized regions and its shift to larger scales with time can be seen in the upper panels of Fig. \ref{fig_ion_neutral_bubbledistr} and \ref{fig_ion_neutral_granulometry}, which show the size distributions of the ionized (top panels) and neutral (bottom panels) regions using the mean free path approach as described in \citet{Lin2016, Giri2018} and the granulometry method as described in \citet{Kakiichi2017}, respectively. 

While the size distribution derived from the mean free path approach\footnote{After marking all ionized cells, we select a random ionized cell. Then we walk in a random axis-aligned direction and count the number of cells until we reach a neutral cell. This process is repeated about $10^4$ times.} reflects the probability of finding an ionized region of size $r$, the size distribution from granulometry represents the probability that an ionized region encompasses an ionized sphere with diameter $r$ (see Appendix \ref{app_granulometry} for its derivation). Hence, if all ionized regions were spheres, both methods would yield the same size distribution of ionized regions. However, as soon as ionized regions deviate from being spheres due to overlap/inhomogeneities in the underlying gas density distribution, the largest sphere fitting into an ionized region will underestimate the actual size of the ionized region at least in one direction; the maximum of the size distribution from granulometry shifts to smaller scales compared to the size distribution from the mean free path approach (cf. black points showing the maximum of the  granulometric size distribution with solid black lines marking the maximum of the mean free path approach size distribution in Fig. \ref{fig_ion_neutral_bubbledistr}).
This also means that the more the probability density distributions of the size distributions derived from the mean free path and granulometry approaches agree with each other, the higher is the probability of the ionized regions having similar extensions in all spatial dimensions.

In order to compare these two size distributions for our three reionization scenarios, in Fig. \ref{fig_ion_neutral_bubbledistr} we show not only the size distribution from the mean free path approach throughout reionization (coloured area) but also mark the maxima of the size distributions derived from the mean free path approach ($R_\mathrm{mfp}$, solid black line) and the granulometry method ($R_\mathrm{g}$, black round points). 
In the case of the ionized regions (top panels in Fig. \ref{fig_ion_neutral_bubbledistr}), $R_\mathrm{mfp}$ and $R_\mathrm{g}$ are in agreement because ionized regions are mostly composed of individual bubbles that have not yet start overlapping significantly at \avchi$\gtrsim0.5$. Thus, in the first half of reionization, ionized regions are likely to resemble sphere-like structures, which can also be seen in the ionization maps shown in Fig. 5 in \citet{Seiler2019}. As the ionized regions grow further and enter the overlap phase, the shapes of the ionized regions become more complex and do not resemble spheres any more. This leads to $R_\mathrm{g}$ dropping below $R_\mathrm{mfp}$.

We have also computed the size distributions of the neutral regions during reionization as shown in the bottom panels of Fig. \ref{fig_ion_neutral_bubbledistr} and \ref{fig_ion_neutral_granulometry}. Firstly, in the initial stages of reionization, $R_\mathrm{g}$ and $R_\mathrm{mfp}$ are in better agreement for the ionized than for the neutral regions. This is because neutral regions are not as likely as ionized regions to resemble sphere-like structures. However, $R_\mathrm{g}$ starts to approach $R_\mathrm{mfp}$ in the final stages of reionization where \avchi$\lesssim0.8$. While throughout most of reionization the shapes of the neutral regions remain non-spherical, they become more confined by the surrounding ionized regions as reionization nears completion. 
In an inside-out reionization scenario found in our simulations, this confinement causes neutral filamentary structures to be more likely to disappear first. This results in ``hemmed-in'' neutral islands in under-dense voids dominating near the end of reionization. These neutral islands tend to have similar extensions in all spatial dimensions.

The size distributions and shapes of the ionized and neutral regions determine the characteristics of the $\chi_\mathrm{HI}$ bispectra (amplitude, sign and scale where it flips signs). As we consider scales comparable to the sizes of the ionized regions, we find the bispectra to be negative for the most squeezed, equilateral and slightly stretched triangles ($0\lesssim\cos\theta\lesssim0.5$). On the other hand, the bispectrum is positive for more stretched isosceles and non-isosceles triangles ($\cos\theta\gtrsim0.5$) and slightly squeezed non-isosceles triangles (see e.g. third and fourth row in Fig. \ref{fig_dimlessbispectra_XHI_multiple} for isosceles and non-isosceles triangles, respectively). 
This overall trend is in agreement with \citet{Majumdar2018} and expected from the above description: the negative contributions from ionized regions (representing below-average concentrations in the $\chi_\mathrm{HI}$ field) dominates at triangles that resemble sphere-like structures, while the positive contributions from neutral regions dominates at triangles that resemble plane-like structures. 

We also study the impact of $f_\mathrm{esc}$ on the link between the characteristics seen in the $\chi_\mathrm{HI}$ bispectra and the size distribution of the ionized and neutral regions throughout reionization. While all our reionization models yield comparable reionization histories, they differ in the ionization topologies due to the varying $f_\mathrm{esc}$ prescriptions. Firstly, from Fig. \ref{fig_ion_neutral_bubbledistr} we see that for all reionization models the peak of the size distributions of ionized (neutral) regions $R_\mathrm{ion}$ ($R_\mathrm{neutral}$) shifts to larger (smaller) scales as reionization proceeds. Secondly and more importantly, as has been already shown in \citet{Seiler2019}, ionized and neutral regions are larger for the {\it SFR} than for the {\it ejected} $f_\mathrm{esc}$ model: as can be seen in Fig. \ref{fig_ion_neutral_bubbledistr} and \ref{fig_ion_neutral_granulometry}, the size distribution is shifted to larger scales as the ionizing emissivity becomes more biased, i.e. $f_\mathrm{esc}$ increases with halo mass.
The reason for the increasing size of the neutral regions as the bias of the ionizing emissivity increases lies in the smaller ionized regions around the lower-mass field galaxies, i.e. galaxies not located in the knots of the large-scale structure. Since the ionized regions around these galaxies critically confine the neutral regions, a decrease in their size corresponds to an increase in the sizes of the neutral regions.

We use the discussion above to explicitly link the triangle configurations, $\chi_\mathrm{HI}$ bispectrum characteristics and sizes of  ionized and neutral regions. We start by reminding the reader that the wave vector $k=\pi/R$ describes features of length $R$ (see Fig. 2 in \citealt{Watkinson2019} for an explanation). Thus, the bispectra in Fig. \ref{fig_dimlessbispectra_XHI_multiple} with $k_1=k_2=0.4h$ and $0.8h$~Mpc$^{-1}$ reflect features with sizes of $r\sim8h^{-1}$ and $4h^{-1}$~Mpc, respectively. In the following we first discuss isosceles triangles ($k_1=k_2$) and move then to non-isosceles triangles with $k_1=\frac{1}{2}k_2$.

\subsubsection{Isosceles triangles:} 
\begin{figure}
 \includegraphics[width=0.50\textwidth]{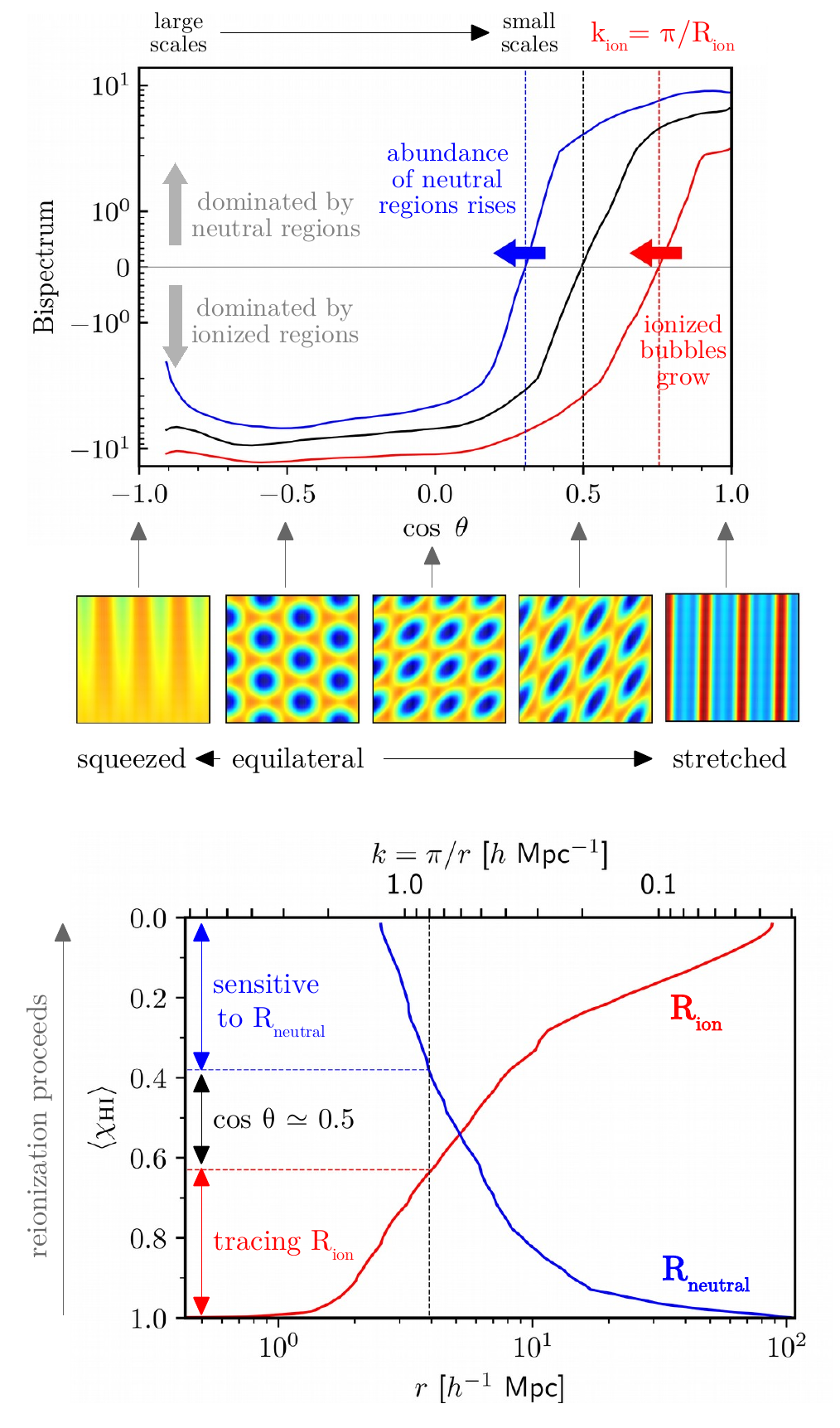}
 \caption{Evolutionary regimes of the $\chi_\mathrm{HI}$ and 21cm bispectrum throughout reionization. The lower panel shows how the three major regimes for isosceles triangles with fixed legs $k_1=k_2=k$ can be determined. In these three regimes the bispectrum is (i) sensitive to the peak of the size distribution of the ionized regions (red), (ii) sensitive to the shapes of the neutral and ionized regions (black), and (iii) sensitive to the abundance and shapes of the neutral regions (blue). The upper panel sketches the typical shape of the bispectra in these regimes, illustrating the typical $\cos\theta$ values at which the bispectra change their sign.}
 \label{fig_interpreting_bispectrum}
\end{figure}

Given the peaks of the typical size distributions of ionized ($R_\mathrm{ion}$) and neutral regions ($R_\mathrm{neutral}$) and the real-space scale of the two triangle legs with fixed lengths in k-space, $r_1=\pi/k_1$ and $r_2=\pi/k_2$, we identify three consecutive regimes within which either the ionized regions, their shapes, or the neutral regions govern the bispectra characteristics:
The lower panel in Fig. \ref{fig_interpreting_bispectrum} also illustrates our definition of these regimes, while the upper panel shows the typical shapes of the bispectra in these regimes, particularly the triangle configurations at which the bispectra switch their signs. 
\begin{table}
  \centering
  \begin{tabular}{c|c|c|c}
    $k$ [$h$ Mpc$^{-1}$] & $\langle \chi_\mathrm{HI} \rangle_\mathrm{min}^\mathrm{f_{ej}}$ & $\langle \chi_\mathrm{HI} \rangle_\mathrm{min}^\mathrm{const}$ & $\langle \chi_\mathrm{HI} \rangle_\mathrm{min}^\mathrm{SFR}$\\
    \hline
    0.3 & 0.3 & 0.4 & 0.5 \\
    0.4 & 0.4 & 0.5 & 0.6 \\
    0.5 & 0.5 & 0.6 & 0.8 \\
    0.6 & 0.7 & 0.8 & 0.9 \\
    0.8 & 0.8 & 0.9 & $>$0.9 
  \end{tabular}
  \caption{Minimum global neutral fraction above which the real-space scale $r=\pi/k$ exceeds the typical size of the ionized regions $R_\mathrm{ion}=\pi/k_\mathrm{ion}$. Values have been inferred from the size distributions of the ionized regions shown in the top row in Fig. \ref{fig_ion_neutral_bubbledistr} and the sign switch of the bispectra for isosceles triangles with values ranging $k_1=k_2=[0.3-1.2]h$~Mpc$^{-1}$.}
  \label{tab_minXHI_ionized_regions}
\end{table}
\begin{table}
  \centering
  \begin{tabular}{c|c|c}
   \hline
   \hline
    $\langle \chi_\mathrm{HI}\rangle$ & $k_\mathrm{ion}^\mathrm{f_{ej}}$ [$h$~Mpc$^{-1}$] & $k_1=k_2$ [$h$~Mpc$^{-1}$]\\
    \hline
    0.6 & 0.8 & 0.4, 0.5\\
    0.7 & 0.9 & 0.5, 0.6\\
    0.8 & 1.1 & 0.5, 0.6\\
    0.9 & 1.6 & 0.7, 0.8\\
    \hline
    \hline
    $\langle \chi_\mathrm{HI}\rangle$ & $k_\mathrm{ion}^\mathrm{const}$ [$h$~Mpc$^{-1}$] & $k_1=k_2$ [$h$~Mpc$^{-1}$]\\
    \hline
    0.6 & 0.7 & 0.4\\
    0.7 & 0.7 & 0.4\\
    0.8 & 1.0 & 0.5, 0.6\\
    0.9 & 1.3 & 0.7, 0.8\\
    \hline
    \hline
    $\langle \chi_\mathrm{HI}\rangle$ & $k_\mathrm{ion}^\mathrm{SFR}$ [$h$~Mpc$^{-1}$] & $k_1=k_2$ [$h$~Mpc$^{-1}$]\\
    \hline
    0.6 & 0.5 & 0.3\\
    0.7 & 0.7 & 0.4\\
    0.8 & 0.7 & 0.4\\
    0.9 & 0.9 & 0.5, 0.6
  \end{tabular}
  
  \caption{For \avchi~values corresponding to the beginning of reionization, we show the $k_\mathrm{ion}=k_{3,t}$ values at which the bispectrum switches its sign (becomes negative), tracing the peak of the size distribution of the ionized regions $R_\mathrm{ion}$ for isosceles triangles. $k_1=k_2$ indicate the bispectra scales (separated by commas) for which those values have been found for isosceles triangles with values ranging $k_1=k_2=[0.3-1.2]h$~Mpc$^{-1}$. For smaller $k_1=k_2$ values than listed the required scale $k_\mathrm{ion}=k_{3,t}$ can not be reached, since the maximum value for $k_3$ is given by $k_3=2k_1=2k_2$. For larger $k_1=k_2$ values one of the real-space scale $r=\pi/k$ traced by the triangle legs drops below the peak of the size distribution of the neutral regions $R_\mathrm{neutral}$ at the corresponding \avchi~value.  For smaller \avchi~values the bispectra do not trace the peak of the size distribution of the ionized regions any more, since one of the conditions of regime (i), $r_1=r_2<R_\mathrm{neutral}$, does not hold any more.}
  \label{tab_transitions}
\end{table}

\paragraph{Beginning of reionization ($r_1=r_2 > R_\mathrm{ion}$ and $r_1=r_2 < R_\mathrm{neutral}$):}
In the beginning of reionization, i.e. as long as all real-space scales $r=\pi/k$, traced by the triangle legs probing the bispectrum, are larger than $R_\mathrm{ion}$, the bispectrum tracks the most abundant sizes of the ionized regions, $R_\mathrm{ion}$. Within this regime (regime (i), red lines in Fig. \ref{fig_interpreting_bispectrum}), the scale $r_{3,t}=\pi/k_{3,t}$, where the bispectrum transitions from being positive to negative for stretched triangles ($k_{3,t}=\sqrt{k_1^2+k_2^2+2k_1k_2\cos\theta_\mathrm{t}}$), corresponds to $R_\mathrm{ion}$ (see red line in Fig. \ref{fig_interpreting_bispectrum}). In order to verify this relation, we have computed the $\chi_\mathrm{HI}$ bispectra for $k_1=k_2$ values ranging from $0.3h$ to $1.2h$~Mpc$^{-1}$. 
Table \ref{tab_minXHI_ionized_regions} lists the neutral fractions above which $r>R_\mathrm{ion}$ holds, while Table \ref{tab_transitions} specifies for which $k_1=k_2$ values the scale $k_{3,t}$ where the bispectrum switches its sign corresponds to $k_\mathrm{ion}=\pi/R_\mathrm{ion}$.
At larger scales, i.e. $k_3<k_{3,t}$, the bispectrum samples structures that correspond to or exceed the peak of the size distribution of the ionized regions; the negative contribution from the ionized regions dominates the bispectrum. 
For smaller scales, i.e. $k_3>k_{3,t}$, the bispectrum samples more plane-like shapes and thus is more sensitive to neutral regions, leading to a positive bispectrum amplitude.
Hence, in the beginning of reionization, when ionized regions are small compared to chosen bispectrum scale $r$, the scale at which the bispectrum switches its sign provides a tracer for the typical size of the ionized regions.
For example, in Fig. \ref{fig_dimlessbispectra_XHI_multiple} this regime applies to the $k_1=k_2=0.4h$~Mpc$^{-1}$ bispectra (second row) at \avchi$\gtrsim0.6$, $0.7$, $0.7-0.8$ for the {\it ejected}, {\it constant} and {\it SFR} $f_\mathrm{esc}$ models, respectively (see black to purple solid lines).

We note that this regime can only be seen for $k_1=k_2$ values that do not exceed the peak of the size distribution of the neutral regions and where the maximum value of $k_3=2k_1=2k_2$ exceeds $k_\mathrm{ion} = \pi/R_\mathrm{ion}$ (c.f. Table \ref{tab_transitions}).

\paragraph{Intermediate stages of reionization ($r_1=r_2 < R_\mathrm{ion}$ and $r_1=r_2 < R_\mathrm{neutral}$):}
As reionization continues, at least one of the real-space scales $r=\pi/k$, traced by the triangle legs, exceeds the peak of the size distribution of the ionized {\it and} neutral regions. In this regime (regime (ii), black lines in Fig. \ref{fig_interpreting_bispectrum}), the bispectrum probes structures with cross sections that are smaller than the peaks of the typical size distributions of ionized or neutral regions. We find the bispectrum to transition from negative to positive values consistently around $\cos\theta_\mathrm{t}\simeq0.5$. This regime can be seen for isosceles triangles probing small to intermediate scales such as for $k_1=k_2=0.8h$~Mpc$^{-1}$ (third row) in Fig. \ref{fig_dimlessbispectra_XHI_multiple}. It lasts longest for the {\it SFR} $f_\mathrm{esc}$ model, \avchi$\simeq0.8-0.3$. As we move to $f_\mathrm{esc}$ models with less biased ionizing emissivities, its duration shortens, e.g. \avchi$\simeq0.85-0.6$ for the {\it constant} $f_\mathrm{esc}$ model, or even vanishes as for the {\it ejected} $f_\mathrm{esc}$ model. This trend can be explained by the smaller values of the peaks of the size distributions of the ionized and neutral regions (see also bottom panel in Fig. \ref{fig_interpreting_bispectrum}).

In this regime, the bispectrum switches its sign at $\cos\theta_t\simeq0.5$, irrespective of the stage of reionization $\langle\chi_\mathrm{HI}\rangle$ or the lengths of the triangle legs $k_1=k_2$.
This independence implies that the triangle configuration at which the bispectrum switches its sign is determined by the dominant shapes of the ionized and neutral regions, primarily by those regions that are smaller than any of the real-space scales $r$ traced by the triangle legs.
Indeed, qualitatively, in the intermediate stages of reionization, before ionized regions significantly overlap, they are more likely to resemble spheres than neutral regions, leading to negative values for $\cos\theta<\cos\theta_t$ and positive values for $\cos\theta>\cos\theta_t$.

We note that this regime only holds as long as the scales probed by the bispectrum do not exceed the scale where the typical ionized and neutral regions equal in size as sketched in the lower panel of Fig. \ref{fig_interpreting_bispectrum}. If this condition fails, we enter a different intermediate regime, in which the evolution of the bispectrum no longer exhibits the same characteristics. We will examine this regime in future work.

\paragraph{End of reionization ($r_1=r_2 < R_\mathrm{ion}$ and $r_1=r_2 > R_\mathrm{neutral}$):}
Towards the end of reionization all real-space scales $r=\pi/k$, traced by the triangle legs, exceed the peak of the size distribution of the neutral regions but not that of ionized regions (regime (iii), blue lines in Fig. \ref{fig_interpreting_bispectrum}). We find the bispectrum to transition from positive to negative values at less and less stretched triangles as reionization proceeds ($\cos\theta_\mathrm{t}<0.5$). This regime can be best seen for $k_1=k_2=0.8h$~Mpc$^{-1}$ (third row) in Fig. \ref{fig_dimlessbispectra_XHI_multiple}: the {\it ejected} $f_\mathrm{esc}$ model enters this regime first around \avchi$\simeq0.7$, followed by the {\it constant} $f_\mathrm{esc}$ model at \avchi$\simeq0.6$, and the {\it SFR} $f_\mathrm{esc}$ model transitions into this regime around \avchi$\simeq0.3$. We also find this regime to prevail when probing larger scales such as $k_1=k_2=0.4h$~Mpc$^{-1}$ (second row) in Fig. \ref{fig_dimlessbispectra_XHI_multiple}, i.e. for \avchi$\lesssim0.5$, $0.5$ and $0.4$ for the {\it SFR}, {\it constant} and {\it ejected} $f_\mathrm{esc}$ models\footnote{If the scales probed by the bispectrum exceed the scale where the peaks of the size distributions of the ionized and neutral regions equal in size, an altered intermediate regime appear where $\pi/k<R_\mathrm{ion}$ and $\pi/k<R_\mathrm{neutral}$. For $k_1=k_2=0.4h$~Mpc$^{-1}$ this applies for the {\it ejected} and {\it constant} $f_\mathrm{esc}$ model. In this regime the bispectrum is sensitive to the typical size and shape of both, ionized and neutral, regions. Consequently, the conditions when this regime is entered and exited change, i.e. it is entered as $\pi/k$ exceeds $R_\mathrm{neutral}$ and exits as $\pi/k$ exceeds $R_\mathrm{ion}$.}.
The shift of the triangle configuration at which the bispectrum switches its sign, $\cos\theta_t$, to less stretched / equilateral triangles as reionization nears completion can be explained as follows: the bispectrum is more sensitive to spherically-shaped regions around the equilateral triangle configuration. As \avchi~decreases, the merged ionized regions continue to grow confining the neutral regions more and more. In an inside-out reionization scenario neutral plane-like structure gradually breaks up into more filamentary structure, and ultimately towards the end of reionization, only isolated neutral islands are left. 

\subsubsection{Non-isosceles triangles:}
Similar regimes to those identified as interesting for isosceles triangles ($k_1=k_2$, $-1\leq\cos\theta\leq1$) can partially also be seen for non-isosceles triangles ($k_1=\frac{1}{2}k_2$). However, due to the different length of the triangles legs $k_1$ and $k_2$, the regime at the intermediate stages of reionization will be altered, since one of the triangle legs will always exceed $R_\mathrm{ion}$ or $R_\mathrm{neutral}$. Hence, in our bispectra plots for non-isosceles triangles (last row in Fig. \ref{fig_dimlessbispectra_XHI_multiple}), we can only clearly identify regime (i) where the scale $r=\pi/k$, at which the bispectrum switches its sign, indicates the peak of the size distribution of the ionized regions $R_\mathrm{ion}$. 

{\it Stretched limit:}
In the stretched limit where the shortest scale of the real-space feature probed by the bispectrum is given by $k_3$ ($\cos\theta>-0.25$ for $k_1=\frac{1}{2}k_2$), regime (i) applies only as long as all real-space scales $r=\pi/k$ traced by the triangle legs exceed $R_\mathrm{ion}$.
As reionization progresses, the bispectrum scales $r_1$ and $r_2$ surpass $R_\mathrm{ion}$ and $R_\mathrm{neutral}$ at different times. During this transition regime and until both scales have exceeded $R_\mathrm{neutral}$ the position of the sign change in the bispectrum is highly dependent on the number of neutral and ionized regions probed. As the more neutral regions fall within the real-space feature probed by the bispectrum, the more positive  the bispectrum becomes. Indeed, for a given \avchi~value, we find that the positions where the bispectrum switches its sign extend  to smaller $\cos\theta$ values for the {\it ejected} than for the {\it SFR} $f_\mathrm{esc}$ model. This aligns with our finding that the {\it ejected} $f_\mathrm{esc}$ model has on average smaller ionized regions than the {\it SFR} $f_\mathrm{esc}$ model. Fixing the length of the triangle legs, the corresponding bispectrum will be more positive for the {\it ejected} than for the {\it SFR} $f_\mathrm{esc}$ model, which is due to the higher abundance of neutral regions smaller than the real-space scales traced.

{\it Squeezed limit:} 
In the squeezed limit, the positive amplitude of the bispectrum for non-isosceles triangles extends to larger $\cos\theta$ values than for isosceles triangles.
The key difference is that the longest scale of the real-space feature for non-isosceles triangles is still in the size range of ionized regions. As we start from a fully squeezed triangle and increase $\cos\theta$, the bispectrum of non-isosceles triangles will probe underlying filamentary structure sooner than a corresponding bispectrum of isosceles triangles.
Hence, the bispectrum tracks the peak of the size distribution of the ionized regions, $R_\mathrm{ion}$, for stretched non-isosceles triangles. 
Considering that for these triangles the shortest scale of the real-space feature is not given by $k_3=\sqrt{k_1^2+k_2^2+2k_1k_2\cos\theta}$ anymore but by the fixed triangle leg $k_2$, the latter starts tracing the typical size of the ionized regions. However, since $k_2$ is fixed, we can only determine the reionization state $\langle\chi_\mathrm{HI}\rangle$ at which the typical size of the ionized regions surpasses the shortest among the bispectrum scales, $r_2=\pi/k_2$. The corresponding characteristic in the bispectrum is the transition from negative to positive values.
Indeed, in the bottom row of Fig. \ref{fig_dimlessbispectra_XHI_multiple} we see that for $k_1=\frac{1}{2}k_2=0.4h$Mpc$^{-1}$ the reionization state, where $k_2$ traces the typical size of the ionized regions, shifts to earlier stages of reionization as the size distribution of the ionized regions becomes more biased and hence its peak shifted to larger scales.
To quantify, the bispectrum becomes positive at $\langle\chi_\mathrm{HI}\rangle\simeq0.7$, $0.8$, $0.9$ for the {\it ejected}, {\it constant} and {\it SFR} $f_\mathrm{esc}$ models, respectively. 
By comparison, for isosceles triangles with $k_1=k_2=0.4h$~Mpc$^{-1}$, this change in sign occurs at larger scales and far later in reionization at $\langle\chi_\mathrm{HI}\rangle\simeq0.3$, $0.2$, $0.2$, respectively.
This shift of the sign change towards later times (lower $\langle \chi_\mathrm{HI} \rangle$ values) as the squeezed bispectrum traces larger scales ($k_2$) has also been found in \citet{Giri2019}, who uses position-dependent power spectra to probe the 21cm bispectra in the squeezed limit.

\subsection{Dependence of the bispectrum on $f_\mathrm{esc}$}

We now explore the signatures of our different $f_\mathrm{esc}$ models that are imprinted in the bispectra throughout reionization.

Firstly, we find that the bispectra of the {\it SFR} $f_\mathrm{esc}$ model show less positive amplitudes at the later stages of reionization  than those of the {\it ejected} and {\it constant} $f_\mathrm{esc}$ models. This trend (similar to the described effect in regime (iii) for isosceles triangles) arises from the different sizes of neutral regions at the given reionization states \avchi: as seen from Fig. \ref{fig_dimlessbispectra_XHI_multiple} the real-space feature described by a fixed triangle samples a larger number of neutral regions when their size distribution is shifted to smaller sizes, leading to a stronger positive contribution to the bispectrum. 
Indeed, as can be seen from Fig. \ref{fig_dimlessbispectra_XHI_multiple} and \ref{fig_ion_neutral_bubbledistr}, the {\it ejected} $f_\mathrm{esc}$ model has smaller neutral regions and more positive bispectra amplitudes at fixed \avchi~than the {\it constant} and {\it SFR} $f_\mathrm{esc}$ models. This effect is noticeable on scales comparable to the sizes of the neutral regions, particularly at the later stages of reionization where \avchi$\simeq0.1-0.2$ (yellow and orange lines in rows 2-4 in Fig. \ref{fig_dimlessbispectra_XHI_multiple}). 

Secondly, comparing the evolution of the $\chi_\mathrm{HI}$ bispectrum during reionization, we see that for fixed $k_1$ and $k_2$ values the range of transition scales where the bispectrum switches its sign becomes smaller as the ionizing emissivity becomes more biased (from the {\it ejected} to the {\it SFR} $f_\mathrm{esc}$ models). This trend is well seen for $k_1=k_2=0.8h$~Mpc$^{-1}$ in the third row of Fig. \ref{fig_dimlessbispectra_XHI_multiple} and can be explained by the evolution of the size distribution of the ionized and neutral regions: as the bias of the ionizing emissivity increases, the size distribution of the ionized and neutral regions shifts to larger sizes and becomes flatter (cf. Fig. \ref{fig_ion_neutral_granulometry}). During the initial stages of reionization, the typical ionized bubble size $R_\mathrm{ion}$ is traced by the bispectrum as long as the chosen triangle corresponds to a real-space feature that comprises this bubble size (see bottom panel in Fig. \ref{fig_interpreting_bispectrum}). Hence, more biased ionizing emissivity models exit this bubble-tracing regime [regime (i)] at an earlier reionization stage. 
Subsequently they enter earlier and stay longer in the regime where the bispectrum is predominantly sensitive to the typical shapes of the ionized and neutral regions [regime (ii)]. 
The reason for the delayed transition from regime (ii) to (iii) lies in the on average larger sizes of the neutral regions, to which the bispectrum becomes then only sensitive in the very last stages of reionization.
Hence, tracking the triangle configuration at which the bispectrum switches its sign, $\cos\theta_t$, can be used to identify these different regimes and thereby gain insight into the size distribution of the ionized and neutral regions during reionization. An ionization topology with larger ionized and neutral regions will remain longer in regime (ii) where $\cos\theta_t$ remains quite constant.

% ***************************************************************************************************
\section{Bispectra of the 21cm signal}
\label{sec_21cm_bispectra}
% ***************************************************************************************************

\begin{figure*}
 \includegraphics[width=0.85\textwidth]{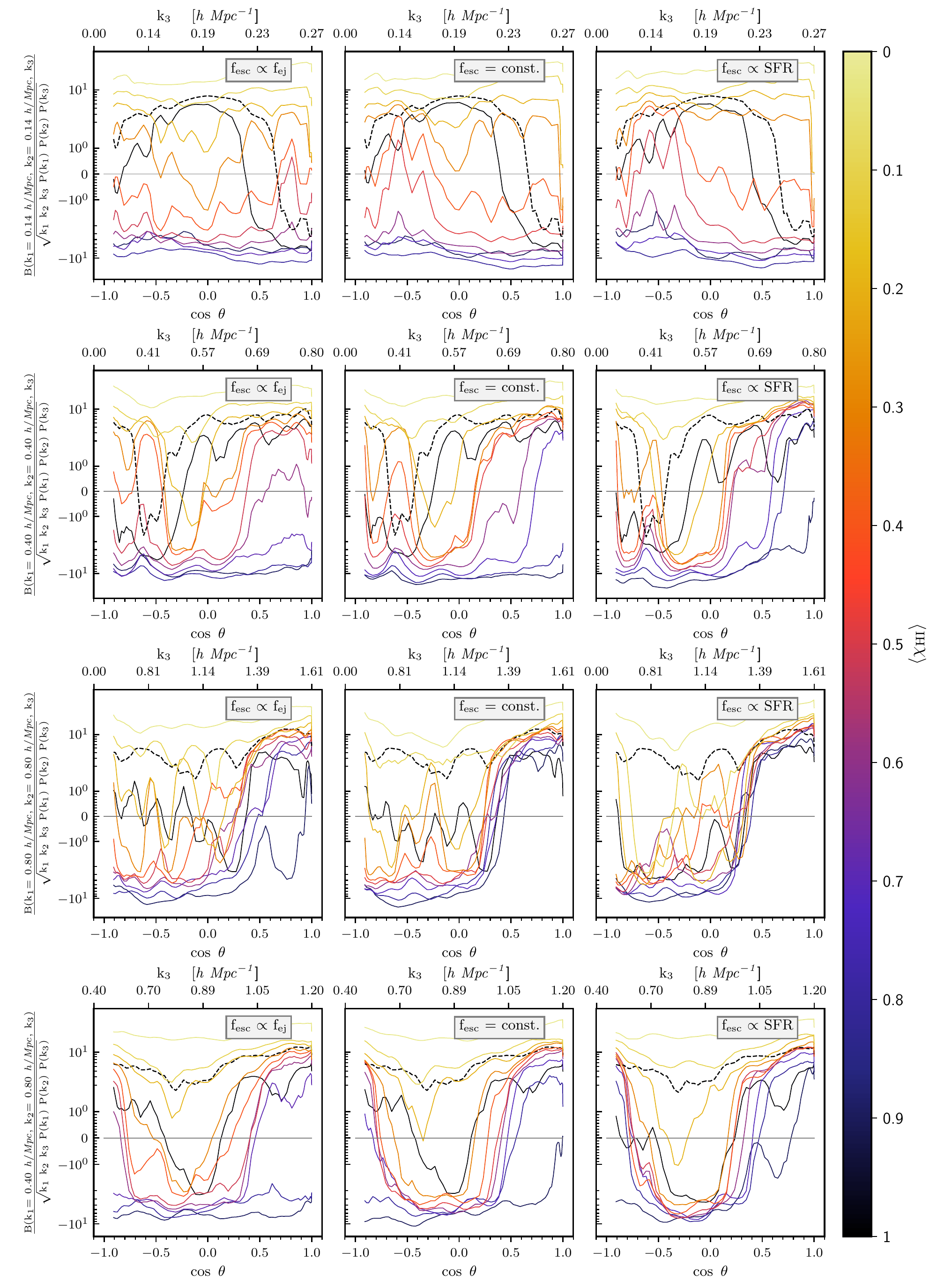}
 \caption{Normalised bispectra of the neutral density ($\delta T_b/T_0$) fluctuations at $\langle \chi_\mathrm{HI} \rangle=0.02$, $0.1$, $0.2$, $0.3$, $0.4$, $0.5$, $0.6$, $0.7$, $0.8$, $0.9$, $0.99$ as indicated by the coloured lines. The black dashed line indicates the normalised bispectrum of the underlying dark matter density field at $\langle \chi_\mathrm{HI} \rangle=0.5$. The left, centre and right columns show the normalised bispectra for the $f_\mathrm{esc}$ models that decrease, remain constant, and increase with halo mass, respectively. The first three rows show the normalised bispectra for isosceles triangles with $k_1=k_2=0.14h$~Mpc$^{-1}$ (probing large scales of $\sim20h^{-1}$Mpc), $0.4h$~Mpc$^{-1}$ (probing intermediate scales of $\sim8h^{-1}$Mpc), $0.8h$~Mpc$^{-1}$ (probing small scales of $\sim4h^{-1}$Mpc), and the last row the normalised bispectra for non-isosceles triangles with $k_1=\frac{1}{2}k_2=0.4h$~Mpc$^{-1}$.}
 \label{fig_dimlessbispectra_21cm_multiple}
\end{figure*}

We derive the normalised bispectra of the 21cm signal $\delta T_b$ by computing the normalised bispectra of $\delta T_b/T_0$, which we show in Fig. \ref{fig_dimlessbispectra_21cm_multiple}.
The 21cm bispectrum depends both on the $\chi_\mathrm{HI}$ and gas density fluctuations. In this Section we focus on the key differences between the 21cm bispectrum and the bispectrum of the $\chi_\mathrm{HI}$ fluctuations, which we have discussed in Section \ref{sec_HI_bispectra}.

Overall the 21cm signal bispectrum follows the $\chi_\mathrm{HI}$ bispectrum closely as long as the non-Gaussianities are dominated by ionized and neutral regions. This applies nearly throughout, except in the very beginning of reionization, when the Universe is almost neutral (\avchi$\gtrsim0.9$) and the only ionized regions appear around the most biased massive galaxies. During those early times the 21cm signal bispectrum is strongly governed by the non-Gaussianities of the underlying gas density. For example, we find the 21cm bispectrum for \avchi$=0.99$ to follow closely the density bispectrum but shifted to smaller values. This shift is due to the highest density peaks being continuously erased from the 21cm maps, as the first ionized regions emerge in the over-dense regions where the most massive galaxies at $z\simeq11$ are located; the emergence of ionized regions correspond to below-average concentrations in the 21cm maps, pushing the 21cm bispectrum to smaller values.
As reionization proceeds further, the contribution of the ionization field to the 21cm bispectrum becomes dominant: the 21cm bispectrum gradually converges to the corresponding $\chi_\mathrm{HI}$ bispectrum until it reaches similar negative amplitudes at \avchi$\simeq0.8$.

As the Universe continues to be ionized (\avchi$\lesssim0.8$), we find the 21cm bispectrum to follow the $\chi_\mathrm{HI}$ bispectrum closely. However, the 21cm bispectrum's additional sensitivity to gas fluctuations introduces moderate deviations from the $\chi_\mathrm{HI}$ bispectrum, both on large and small scales.

On larger scales the amplitude of the 21cm bispectrum is slightly shifted to smaller values for \avchi$\gtrsim0.2$. The reason for this shift in amplitude lies in the inside-out character of our reionization simulations. 
Since the ionization fronts percolate from the most over-dense regions into the under-dense regions of the IGM, the largest peaks of the 21cm signal in over-dense knots are erased first. 
While parts of the over-dense filaments in the IGM remain neutral, their fraction in the neutral volume is less than $\sim10\%$ \citep{Hutter2017} and the majority of the neutral volume is under-dense. It is the latter that causes the contribution of above-average concentrations, coming from the over-dense neutral regions, to be lower in the 21cm than in the $\chi_\mathrm{HI}$ fields.
The discrepancy between the 21cm and $\chi_\mathrm{HI}$ fields in terms of contrast, i.e. $\frac{\delta T_b({\bf x})}{\delta T_b^\mathrm{max}}$ and $\frac{\chi_\mathrm{HI}({\bf x})}{\chi_\mathrm{HI}^\mathrm{max}}$, is strongest in voids where the 21cm signal is lower due to the under-density while $\chi_\mathrm{HI}$ is at its maximum. Hence, we expect that the shift in amplitude to smaller values is stronger at triangle combinations whose real-space structures correspond to under-dense voids.
Indeed, from the second rows in Fig. \ref{fig_dimlessbispectra_XHI_multiple} and \ref{fig_dimlessbispectra_21cm_multiple} we see that for $k_1=k_2=0.4h$~Mpc$^{-1}$ and $k_3=0.36h$~Mpc$^{-1}$ the maximum is weaker in the 21cm bispectra than in the $\chi_\mathrm{HI}$ bispectra for all $f_\mathrm{esc}$ models.

On smaller scales, the $\cos\theta$ values, where the 21cm bispectrum changes its sign, are slightly shifted towards equilateral triangles compared to the $\chi_\mathrm{HI}$ bispectrum. These shifts are of similar order for isosceles triangle configurations ($\Delta\cos\theta\lesssim0.1-0.2$), and can be explained as follows. 
As the bispectra probe smaller scales, i.e. for increasing $k_1$ and $k_2$ values, they approach the scales of filaments in the underlying density field while the number of ionized regions, probed by the bispectra, decreases. Hence, the negative contribution from the ionized regions to the 21cm bispectrum is counteracted by an increasing positive contribution from filamentary structure in the underlying density field. This contribution of the gas density field to the 21cm bispectrum causes not only the sign change of the 21cm bispectrum to shift towards equilateral triangles but also the negative amplitude around equilateral to moderately stretched triangles to decrease (cf. third rows in Fig. \ref{fig_dimlessbispectra_XHI_multiple} and \ref{fig_dimlessbispectra_21cm_multiple}). 

Finally, we find our 21cm bispectra to be in good agreement with the findings in \citet{Majumdar2018}. Their model results and our {\it constant} $f_\mathrm{esc}$ model show very similar $\cos\theta$ values where the bispectrum switches sign for $k_1=k_2=0.8h$Mpc$^{-1}$ and $k_1=\frac{1}{2}k_2=0.4h$Mpc$^{-1}$ (cf. Fig. \ref{fig_dimlessbispectra_21cm_multiple} in this paper and Fig. 7 in \citet{Majumdar2018}). Furthermore, we also find the change of sign of the 21cm bispectrum to be shifted by a similar extent towards equilateral triangles compared to the $\chi_\mathrm{HI}$ bispectrum.
However, we note that the negative part of our bispectra show more fluctuations in amplitude than in \citet{Majumdar2018}. On the one hand, this difference may be due to the larger minimum halo mass of the ionizing sources ($M_\mathrm{min}=1.1\times10^9\msun$ compared to $M_\mathrm{min}=4\times10^8\msun$) while having a similar resolution of the underlying density field ($\sim0.6$Mpc), leading to the ionized regions around the least massive halos in the simulation being better resolved. On the other hand, the coarser binning of $k$-values in \citet{Majumdar2018} causes the amplitude fluctuations to be smoothed. Indeed, since a larger number of triangles decreases the noise in the statistic, an increase of our uncertainties in $k$ leads to smoother bispectra. However, it also averages over such large ranges of real-space scales (particularly for large real-space modes) that key large-scale features in the $\chi_\mathrm{HI}$ or 21cm bispectra disappear, which is why we use a finer binning of $k$ values.

% ***************************************************************************************************
\section{Discussion \& Conclusions}
\label{sec_conclusions}
% ***************************************************************************************************

We have analysed the bispectra of the $\chi_\mathrm{HI}$ and 21cm differential brightness temperature fluctuations during the Epoch of Reionization for three different reionization scenarios that combine merger trees from a N-body simulation with a self-consistent semi-numerical model describing galaxy evolution and reionization simultaneously \citep[{\sc rsage},][]{Seiler2019}. For each simulation, we have assumed a different model of the escape fraction of \HI ionizing photons $f_\mathrm{esc}$: (1) $f_\mathrm{esc}$ scales with the fraction of gas ejected from the galaxy, i.e. the $f_\mathrm{esc}$ values for low mass halos are higher than for high mass halos, leading to a more homogeneous ionizing emissivity distribution and a high abundance of small-to-medium-sized ionized regions. (2) $f_\mathrm{esc}$ is constant. (3) $f_\mathrm{esc}$ scales with the SFR of the galaxy, i.e. the $f_\mathrm{esc}$ values increase with halo mass, resulting in a very biased ionizing emissivity distribution and very large (small) ionized regions around the most (least) massive galaxies. 

Our key aim is to link characteristic features in the 21cm bispectrum to (i) the size distribution of the ionized and neutral regions and (ii) the large-scale reionization topology. Our key findings are:
\begin{itemize}
 \setlength\itemsep{0.5em}
 \item The bispectrum of the 21cm signal follows the $\chi_\mathrm{HI}$ bispectrum closely throughout reionization. Only in the very beginning of reionization, \avchi$\gtrsim0.9$, does it deviate towards the bispectrum of the density fluctuations. This trend is in agreement with findings in \citet{Shimabukuro2016} and \citet{Majumdar2018}.
 \item Considering large-scale voids, the 21cm and $\chi_\mathrm{HI}$ bispectra trace are extremely sensitive whether the ionization percolate from over-dense to under-dense regions (inside-out) or vice versa (outside-in). Here, we show that for an inside-out reionization topology, the minima in the density bispectrum correspond to the maxima in the 21cm bispectrum.
 \item From squeezed to stretched triangles, the 21cm bispectra features a change of sign from negative to positive values, where ionized regions representing below-average concentrations contribute negatively and neutral regions representing above-average concentrations positively. Consequently, the position of the change of sign depends strongly on the typical sizes of the ionized and neutral regions, as well as on the scales that are probed with the bispectrum. As long as the corresponding real-space features of a ($k_1$, $k_2$, $k_3$) triangle configuration are larger than the peak of the size distribution of the ionized regions $R_\mathrm{ion}$, the change of sign in the bispectrum occurs at the corresponding $k_i=\pi/R_\mathrm{ion}$ value. This provides a direct tracer of the typical size of the ionized regions during the earlier stages of reionization.
 
 \item The 21cm bispectrum traces the ionization topology, i.e. depends strongly on the evolution of the size distributions of the ionized and neutral regions. For fixed triangle legs $k_1$ and $k_2$ in {\it isosceles} triangles, we find three major regimes for the 21cm and $\chi_\mathrm{HI}$ bispectrum: 
 \begin{enumerate}
  \setlength\itemsep{0.5em}
  \item The corresponding real-space features of a ($k_1$, $k_2$, $k_3$) triangle configuration is larger (smaller) than the peak of the size distribution of the ionized (neutral) regions $R_\mathrm{ion}$. In this case the scale at which the bispectrum changes its sign ($k_3$) traces the peak of the size distribution of the ionized regions ($R_\mathrm{ion}$).
  \item The corresponding real-space features of a ($k_1$, $k_2$, $k_3$) triangle configuration is smaller than the peak of the size distribution of the ionized and neutral regions. In this case the change of sign in the bispectrum stagnates around values of $\cos\theta_t\simeq0.5$ and $\cos\theta_t\simeq0.3$ for $\chi_\mathrm{HI}$ and 21cm fields, respectively.
  \item The corresponding real-space features of a ($k_1$, $k_2$, $k_3$) triangle configuration is larger (smaller) than the peak of the size distribution of the neutral (ionized) regions $R_\mathrm{ion}$. In this case the scale at which the bispectrum changes its sign ($k_3$) moves towards equilateral triangles during the end stages of reionization as the abundance of confined neutral regions increases.
 \end{enumerate}
 The \avchi~values at which the bispectrum ``transitions'' into a new regime depends strongly on the underlying size distribution of the ionized and neutral regions and the bispectrum scales being considered. A more biased ionizing emissivity, such as our {\it SFR} $f_\mathrm{esc}$ model, leads to a flatter size distribution of the ionized and neutral regions that is shifted to larger scales. Hence, for the same ($k_1$, $k_2$) isosceles triangles, a size distribution of ionized regions that is shifted to larger scales leads to an earlier exit (at higher \avchi~values) from regime (i) and entry into regime (ii). Furthermore, it exits regime (ii) and enters regime (iii) at a later time (lower \avchi~values), since its neutral regions are on average larger.
 Hence, by identifying at which \avchi~values the 21cm bispectrum enters and exists regime (ii) we can pinpoint the peaks of the size distributions of the ionized and neutral regions at these times, respectively. Decreasing the length of the triangle legs $k_1$ and $k_2$, i.e. probing larger real-space fluctuations, decreases (increases) the \avchi~value when the 21cm bispectrum enters (leaves) regime (ii), and would allow to systematically probe the typical size of the ionized and neutral regions throughout the epoch of reionization.
 
 \item The 21cm bispectrum for squeezed non-isosceles triangles traces the peak of the size distribution of the ionized regions. It changes its sign from negative to positive values when the peak of the size distribution of the ionized regions surpasses the real-space scale of the longest triangle leg ($k_1$ or $k_2$). Hence, measuring the 21cm bispectra of squeezed non-isosceles triangles for a range of $k_1=\frac{1}{n}k_2$ values would enable observers to trace the growth of the ionized regions.
\end{itemize}

These findings highlight that the 21cm bispectrum provides a valuable tracer of the ionization topology and gives more detailed insight into the size distribution of the ionized regions during reionization than the 21cm power spectrum.

We end by summarizing the major caveats of this work. Firstly, the limited simulation box size of $160$~Mpc causes the bispectra probing larger scales to be subject to statistical variance. The low number of triangles could be increased by increasing the uncertainty of the $k_1$, $k_2$, $k_3$ values, i.e. their bin width. However, with this uncertainty being determined by our simulation grid, this would result in unreasonable high uncertainties corresponding to $\Delta r\sim6-20$~Mpc in real-space.

Secondly, our 21cm signal modelling assumes the spin temperature to be heated above the CMB throughout reionization. This approximation possibly breaks down at the very early stages of reionization, making the 21cm signal also subject to spin temperature fluctuations. Accounting for spin temperature fluctuations would lead to a slight decrease in the 21cm bispectra amplitudes due to a lower 21cm signal in the under-dense regions. This would result in a little change in the qualitative shape of the 21cm bispectra. 

Thirdly, all our reionization simulations show an inside-out reionization topology, where the ionization fronts originate in the over-dense regions and percolate into the under-dense regions last. Deriving the 21cm bispectra for an outside-in reionization scenario would  confirm our interpretation of the sensitivity of the 21cm bispectrum to the reionization topology.

We end by noting that, for some wave modes and triangle configurations, the 21cm bispectrum may require less observing time than the 21cm power spectrum \citep{Trott2019}. Hence, in addition to the 21cm power spectrum, the 21cm bispectrum provides a powerful tool to place further constraints on astrophysical parameters from forthcoming 21cm signal detections. It will be key in constraining the size distribution of ionized regions and the large-scale reionization topology, which again, together with high-redshift galaxy observations, will provide crucial insights into the nature of the ionizing sources in our Universe.

% ***************************************************************************
\section*{Acknowledgements} 
% ****************************************************************************
AH and PD acknowledge support from the European Research Council's starting grant ERC StG-717001. JS and AH have been supported under the Australian Research Council’s Discovery Project funding scheme (project number DP150102987). Parts of this research were
conducted by the Australian Research Council Centre of Excellence for All Sky Astrophysics in 3 Dimensions (ASTRO 3D), through project number CE170100013.
CAW acknowledges financial support from the European Research Council under ERC grant number 638743-FIRSTDAWN (held by Jonathan Pritchard).

% **************************************************************************
% REFERENCES
% **************************************************************************

\bibliographystyle{mn2e}
\bibliography{thisBib}

% **************************************************************************
% APPENDIX
% **************************************************************************

\appendix

% ***************************************************************************
\section{Granulometry}
\label{app_granulometry}
% ***************************************************************************

\begin{figure*}
 \includegraphics[width=0.8\textwidth]{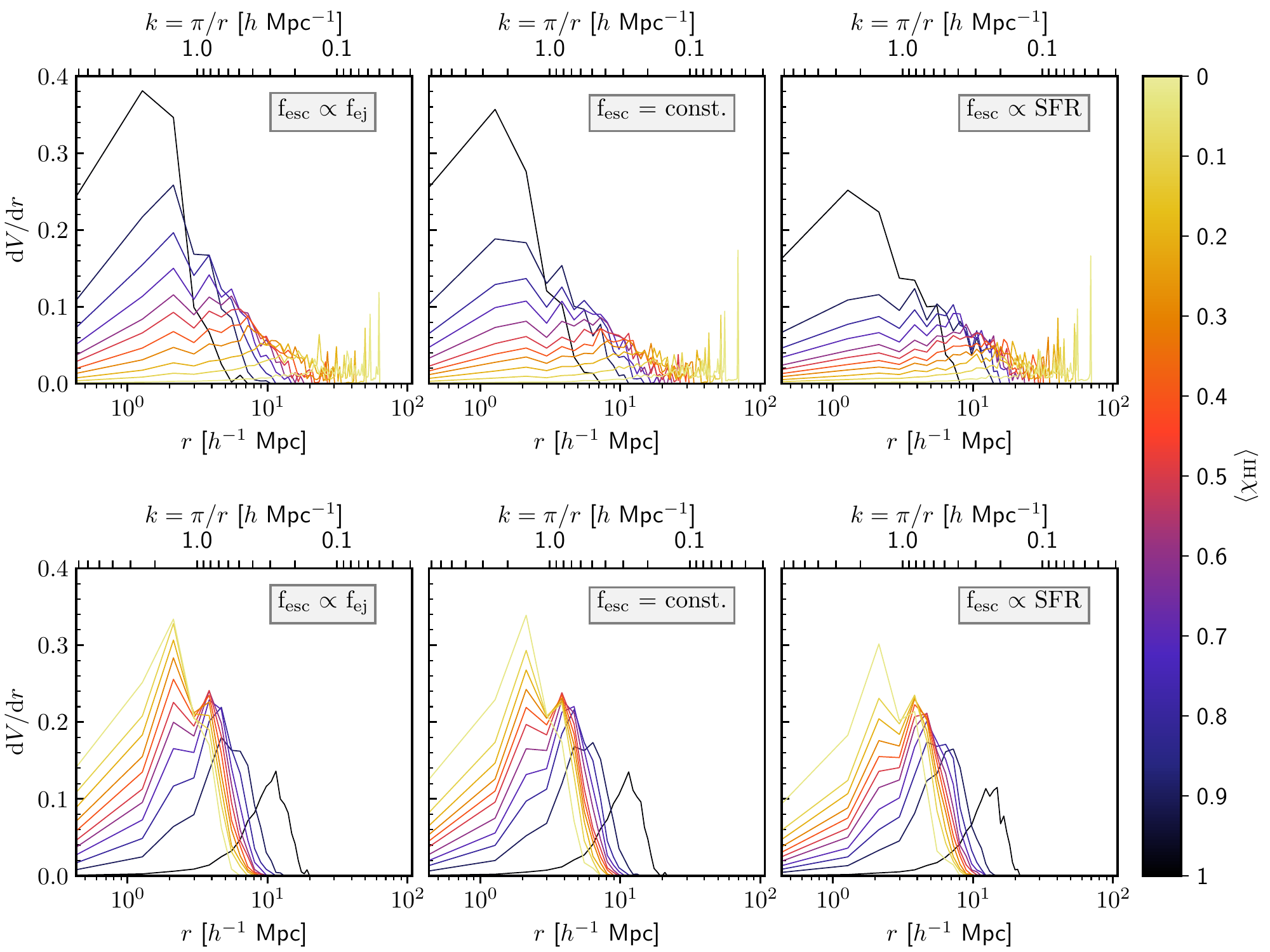}
 \caption{Top: Granulometry for the ionized regions. Bottom: Granulometry for the neutral regions.}
 \label{fig_ion_neutral_granulometry}
\end{figure*}

We compute the granulometry of the ionization fields $F$ by employing the method outlined in \citet{Kakiichi2017}. Assuming for the structuring element $S$ a sphere with diameter $r$, we determine the ionized (neutral) volume $V(r)$ that can be composed of such spheres by using the morphological opening operation
\begin{eqnarray}
 F \circ S = (F \ominus S) \oplus S.
\end{eqnarray}
We determine the volume $V(r)$ for varying sphere sizes $r$, and construct the size distribution of ionized (neutral) spheres by computing the derivative $\mathrm{d}V/\mathrm{d}r$. The resulting size distributions are shown throughout reionization for our three $f_\mathrm{esc}$ models ({\it ejected}, {\it constant}, {\it SFR}) in Fig. \ref{fig_ion_neutral_granulometry}.

% ***************************************************************************
\section{Statistical Fluctuations}
\label{app_stats_fluctuations}
% ***************************************************************************

\begin{figure*}
 \includegraphics[width=1.02\textwidth]{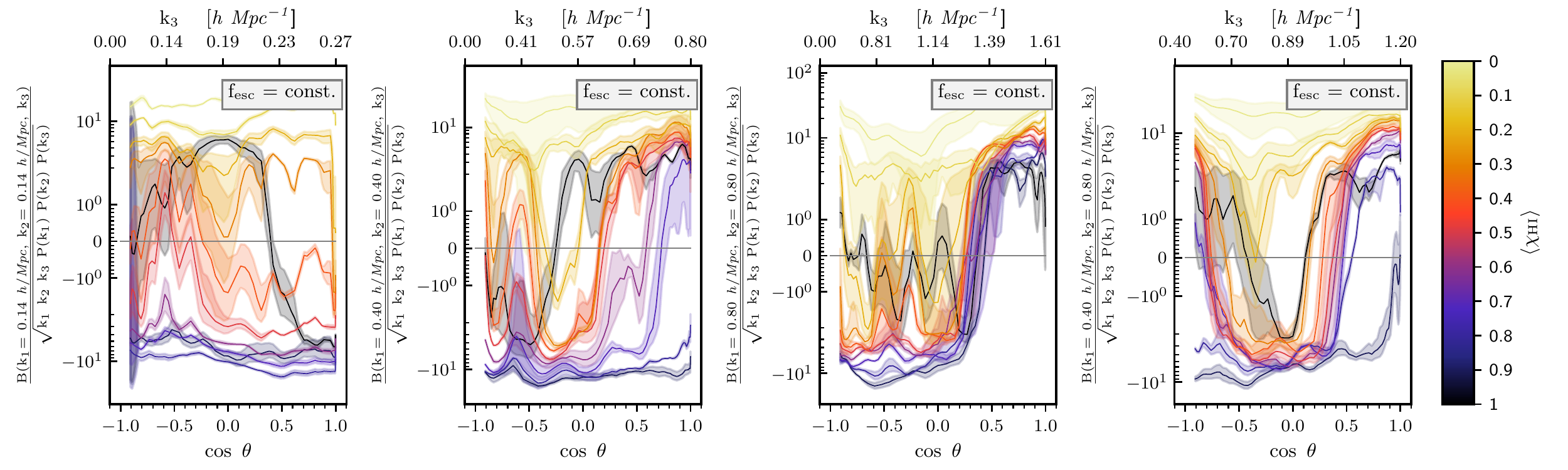}
 \caption{Normalised bispectra of the neutral density ($\delta T_b/T_0$) fluctuations at $\langle \chi_\mathrm{HI} \rangle=0.02$, $0.1$, $0.2$, $0.3$, $0.4$, $0.5$, $0.6$, $0.7$, $0.8$, $0.9$, $0.99$ as indicated by the coloured lines, for the {\it constant} $f_\mathrm{esc}$ model. Transparent shaded regions show the corresponding uncertainties derived using the Jackknife resampling. From left to right the panels show the normalised bispectra for isosceles triangles with $k_1=k_2=0.14h$~Mpc$^{-1}$ (probing large scales of $\sim20h^{-1}$Mpc), $0.4h$~Mpc$^{-1}$ (probing intermediate scales of $\sim8h^{-1}$Mpc), $0.8h$~Mpc$^{-1}$ (probing small scales of $\sim4h^{-1}$Mpc), and the normalised bispectra for non-isosceles triangles with $k_1=\frac{1}{2}k_2=0.4h$~Mpc$^{-1}$.}
 \label{fig_dimlessbispectra_21cm_stats_error}
\end{figure*}

In order to estimate the uncertainties in our bispectra due to statistical fluctuations, we use the Jackknife resampling method \citep{Norberg2009, Wu1986}. For this purpose, we divide our simulation box into $n=8$ subboxes and compute the bispectrum $B_i$ for each subbox $i$. The mean bispectrum of all subboxes is then given by
\begin{eqnarray}
 \langle B \rangle &=& \frac{1}{n} \sum_{i=1}^n B_i.
\end{eqnarray}
To build the i-th Jackknife replication, we calculate the mean bispectra of all subboxes except subbox $i$.
\begin{eqnarray}
 \langle B_i \rangle &=& \frac{1}{n-1} \sum_{j=1, j\neq i}^n B_j
\end{eqnarray}
The Jackknife estimate of the standard error is defined as
\begin{eqnarray}
 \sigma_B &=& \left( \frac{n-1}{n} \sum_{i=1}^n \left[ \langle B_i \rangle - \langle B \rangle \right]^2 \right)^{1/2}.
\end{eqnarray}
We employ the same method to estimate the Jackknife standard error for the normalised bispectrum $\tilde{B}$, where each $\tilde{B}_i$ has been normalised by the powerspectrum $P_i$ of the respective subbox.
We show the bispectrum computed from the entire box and the Jackknife standard error derived from the subboxes for the {\it constant} $f_\mathrm{esc}$ model in Fig. \ref{fig_dimlessbispectra_21cm_stats_error}. From this Figure we can see that the position where the bispectrum switches its sign at stretched (stretched and squeezed) triangles for isosceles (non-isosceles) triangles is well constrained, particularly during reionization. Uncertainties increase as the bispectrum oscillates around zero, as it is the case for $k_1=k_2=0.8h$~Mpc$^{-1}$ at \avchi$\simeq0.99$ and $0.1$.
Furthermore, we find large-scale fluctuations ($k\lesssim0.2h$~Mpc$^{-1}$) to be subject to increased uncertainties. This is not surprising as these scales approach the size of the simulation subbox and the number of triangles probed decreases accordingly.

\end{document}